\documentclass[%
 reprint,
 amsmath,amssymb,
aps,
prd,twocolumn,superscriptaddress,
longbibliography
]{revtex4-1}
\usepackage{braket}
\usepackage{graphicx}
\usepackage{bm}
\usepackage{float}
\usepackage{color} 
\usepackage{soul} 
\usepackage[caption = false]{subfig}
\usepackage{mathbbol}
\usepackage{placeins}
\usepackage[english]{babel}

\newcommand{\kb}{k_\mathrm{B}}

\newcommand{\Ax}{A_1}
\newcommand{\Ay}{A_2}
\newcommand{\Az}{A_3}

\newcommand{\Ix}{I_1}
\newcommand{\Iy}{I_2}
\newcommand{\Iz}{I_3}

\newcommand{\Lz}{L_3}
\newcommand{\Ly}{L_2}
\newcommand{\Lx}{L_1}

\newcommand{\sLz}{{\sf L}_3}
\newcommand{\sLy}{{\sf L}_2}
\newcommand{\sLx}{{\sf L}_1}

\newcommand{\Jo}{J_0}

\usepackage[normalem]{ulem}

\begin{document}

\title{Quantum persistent tennis racket dynamics of nanorotors}

\author{Yue Ma}
\affiliation{QOLS, Blackett Laboratory, Imperial College London, London SW7 2AZ, United Kingdom}

\author{Kiran E. Khosla}
\affiliation{QOLS, Blackett Laboratory, Imperial College London, London SW7 2AZ, United Kingdom}

\author{Benjamin A. Stickler}
\email{b.stickler@imperial.ac.uk}
\affiliation{QOLS, Blackett Laboratory, Imperial College London, London SW7 2AZ, United Kingdom}
\affiliation{Faculty of Physics, University of Duisburg-Essen, 47048 Duisburg, Germany}

\author{M. S. Kim}
\email{m.kim@imperial.ac.uk}
\affiliation{QOLS, Blackett Laboratory, Imperial College London, London SW7 2AZ, United Kingdom}

\begin{abstract}

\noindent \rm{{Classical rotations of asymmetric rigid bodies are unstable around the axis of intermediate moment of inertia, causing a flipping of rotor orientation. This effect, known as the tennis racket effect, quickly averages to zero in classical ensembles since the flipping period varies significantly upon approaching the separatrix. Here, we explore the quantum rotations of rapidly spinning thermal asymmetric nanorotors and show that classically forbidden tunnelling gives rise to persistent tennis racket dynamics, in stark contrast to the classical expectation. We characterise this effect, demonstrating that quantum coherent flipping dynamics can persist even in the regime where millions of angular momentum states are occupied. This persistent flipping offers a promising route for observing and exploiting quantum effects in rotational degrees of freedom for molecules and nanoparticles.}}
\end{abstract}

\maketitle

Quantum control of nanomechanical motion is a challenging task with great potential for future quantum technologies and fundamental tests of physics \cite{aspelmeyer2014cavity}. Optically levitating nanoparticles in ultrahigh vacuum achieves an unrivalled degree of environmental isolation \cite{millen2019optomechanics}, rendering these systems ideally suited for high-precision sensing \cite{ranjit2016zeptonewton,chaste2012nanomechanical,kuhn2017optically} and for high-mass tests of quantum physics \cite{romero2011large,scala2013matter,bateman2014near,stickler2018probing}. In addition, freely suspending nanoscale dielectrics introduces rigid-body rotations as novel nanomechanical degrees of freedom.

The non-linearity and anharmonicity of nanoparticle rotations induces quantum interference effects which have no analogues in their free centre-of-mass motion. The rotations of levitated objects can be controlled and exploited by using aspherical or anisotropic objects rather than isotropic spheres \cite{hoang2016torsional,kuhn2017full}. Amongst the most recent experimental achievements are frequency-locking of nanorods \cite{kuhn2017optically}, GHz rotations of anisotropic spheres \cite{reimann2018} and nanodumbbells \cite{ahn2018optically,ahn2020ultrasensitive}, and rotational precession of dielectrics \cite{rashid2018precession}. Together with the breakthroughs of centre-of-mass ground state cooling of spherical particles \cite{delic2020cooling} {and rotational cooling of aspherical particles \cite{delord2020spin,bang20205d}}, these developments pave the way for cooling the rotational state into the quantum regime \cite{stickler2016rotranslational,seberson2019parametric} and can be used for probing quantum physics \cite{ma2017proposal,stickler2018probing,hummer2020} with nano- to microscale rigid rotors.

\begin{figure}[t]
\centering
\includegraphics[width=0.49\textwidth]{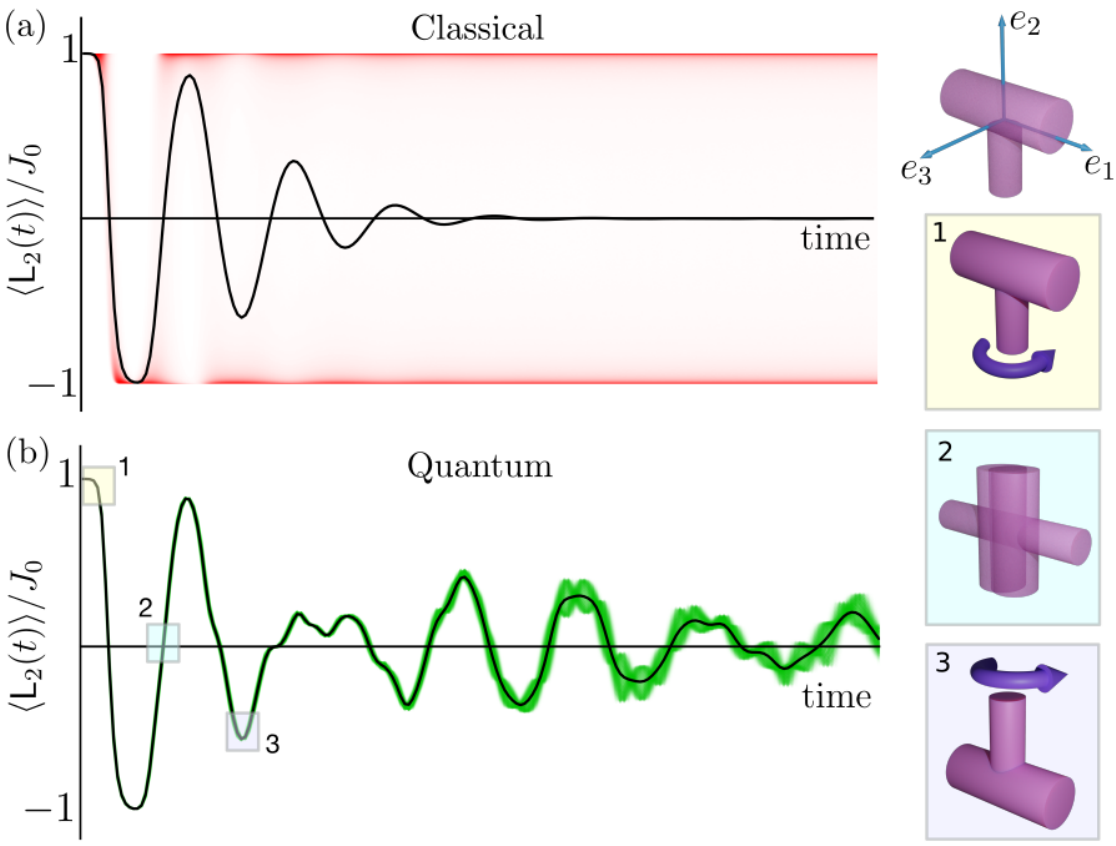}
\caption{An asymmetric rigid rotor initially rotating with angular momentum $\Ly$ around its mid-axis $e_2$, exhibits periodic flipping of its orientation, known as the tennis racket effect. The flipping of its orientation is manifested by a periodic change of sign of the mid-axis angular momentum, $\langle {\sf L}_2(t)\rangle/J_0$. {The red and green shaded regions show the probability density of mid-axis angular momentum values.} (a) {Gaussian decay of mix-axis flipping classically.} (b) {Persistent mid-axis flipping quantum mechanically.} The insets (1-3) show the quantum mid-axis orientation at the times highlighted in (b).}\label{fig:model}
\end{figure}

\begin{figure*}
\centering
\includegraphics[width=1\textwidth]{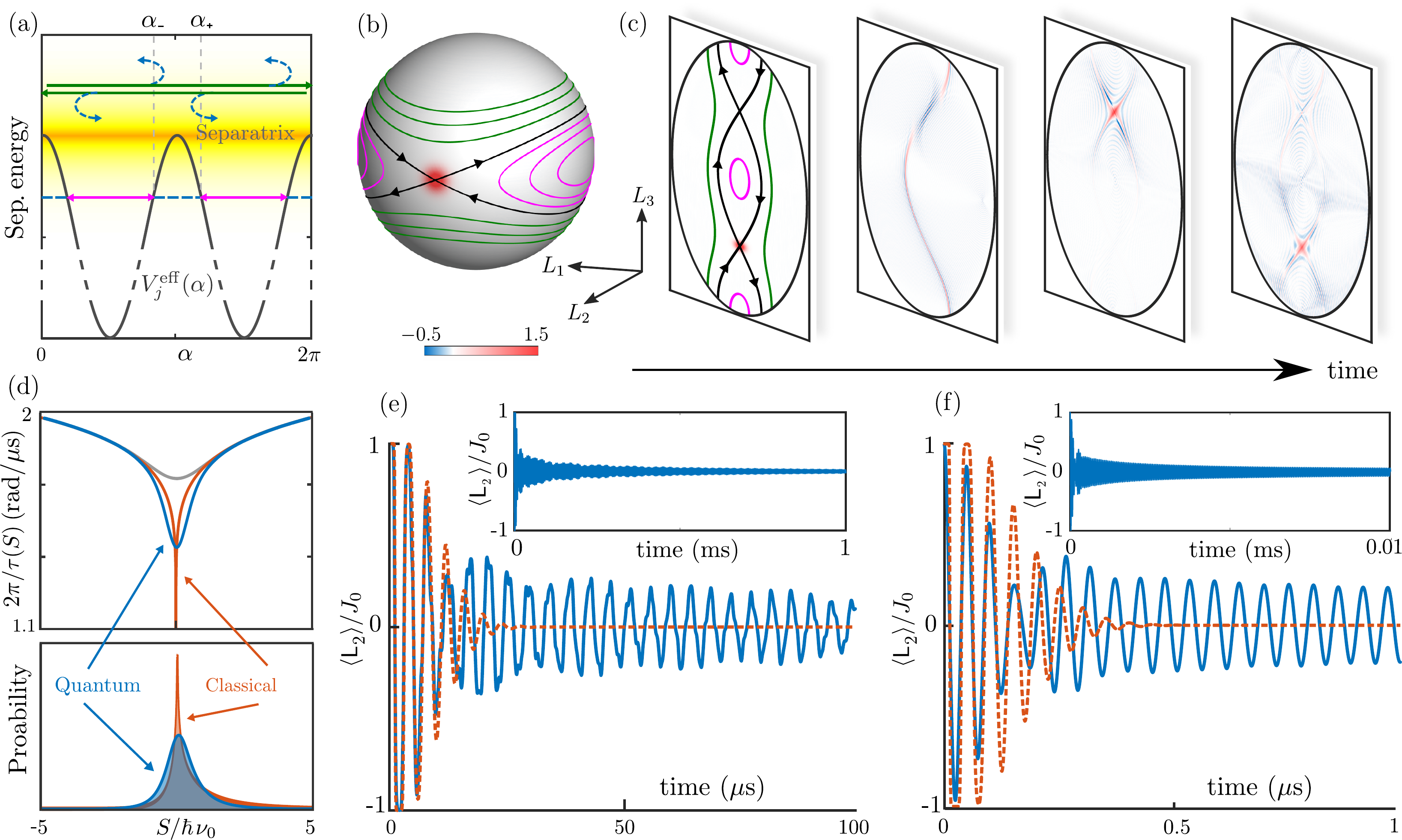}
\caption{(a) The unstable mid-axis dynamics of an asymmetric top is physically similar to that of a planar rotor in a periodic potential (see the main text for details). {The color density shows the probability of different separatrix energies around $S=0$ (Eq.~\eqref{eq:Pr_S}), with dashed arrows indicating quantum reflections above the potential barrier.} (b) {Angular momentum sphere shows the separatrix (black line) between classical rotational (green lines) and oscillatory (pink lines) trajectories for fixed $j=100$.} (c) {Time evolution of the Wigner function shows periodic dispersion and partial interference (negative value in blue).} (d) {Upper (Lower) panel: Quantum and classical frequencies (relative probabilities of energy population) close to the separatrix.}  (e-f): The quantum mid-axis dynamics (blue lines) persist orders of magnitude longer than the corresponding classical dynamics (orange dashed line). (e) Calculated exactly for $\Jo/\hbar = 10^6$ by full numerical diagonalisation. (f) Calculated from the approximation \eqref{eq:quantLapprx} for $\Jo/\hbar = 10^8$, where a full numerical treatment is impossible.}\label{fig:2}
\end{figure*}

The instability of classical rigid body rotations around the axis of intermediate moment of inertia results in an exponential growth of initial deviations from perfect mid-axis revolutions \cite{landau1976mechanics}. The resulting periodic flipping of the orientation of asymmetric rigid bodies, which was first studied by Poinsot in the 19th century~\cite{Poinsot}, is now known as the tennis racket (or Dzhanibekov) effect. {Analogies to this classical flipping effect have been exploited in several fields, for instance, qubit control~\cite{damme_linking_2017}, non-equilibrium quantum many-body simulation~\cite{gerving2012non} and Josephson junction dynamics~\cite{chuchem2010quantum}}. The frequency of the tennis racket flips crucially depends on the initial rotation state, implying that they quickly vanish when averaged over repetitions of the experiment (or in an ensemble of particles), see Fig.~\ref{fig:model}(a). However, in this paper we show that the mid-axis rotations become persistent in the quantum regime, see Fig.~\ref{fig:model}(b).

We argue that quantum-persistent tennis racket flips are realistically observable with levitated nanoparticles at moderate motional temperatures. Such an experiment would probe quantum rotations in an unprecedented mass and complexity regime. Furthermore, we argue that the quantum tennis racket flipping can be understood by a tunnelling-induced splitting of degenerate eigenenergies close to the classical phase space instability. We discuss the quantum and classical rotor dynamics with analytical and numerical methods, demonstrating that quantum persistent flipping can be observable with GHz nanorotors at milliKelvin temperatures. 

\emph{Classical dynamics} -- 
The rotation dynamics of an asymmetric rigid rotor with moments of inertia $\Ix > \Iy > \Iz$ is integrable, conserving the total angular momentum $L^2 = \Lx^2 + \Ly^2 + \Lz^2$ and the energy $H = \Ax\Lx^2 + \Ay \Ly^2 + \Az \Lz^2$, where $A_i =1 /2I_i$ are the rotation constants ($\Az > \Ay > \Ax$). The body-fixed angular momentum trajectories, $L_i(t)$, follow from Euler's equations of motion \cite{landau1976mechanics,hamraoui2018,ashbaugh_twisting_1991}. If the initial rotation is closely aligned to the mid axis, $\Ly \gg \Lx \simeq\Lz$, the rotation dynamics become unstable, resulting in flipping between the initial angular momentum $\Ly$ and the opposite rotor orientation $-\Ly$, \cite{van2017tennis}. In this limit, the body-fixed mid angular momentum simplifies to $\Ly(t) \simeq \Ly \mathrm{cd}\left(\nu t, q\right)$, where ${\rm cd}(\cdot)$ is the Jacobi CD elliptic function, with $\nu = 2\sqrt{(\Ay-\Ax)(\Az - \Ay)}L$ and $q \simeq 1 - 4 (\Az - \Ax) \vert S \vert/\nu^2$ the elliptic modulus. Here the separatrix energy $S = H - \Ay L^2$ quantifies how close the dynamics is to the rotor phase space instability at $S = 0$, which defines the separatrix.

The period of the flipping follows from the properties of the Jacobi functions,
\begin{equation} \label{eq:period}
\tau_{\rm cl}(S) \simeq \frac{2}{\nu} \ln \left [ \frac{4 \nu^2}{(\Az - \Ax) \vert S \vert}\right ],
\end{equation}
and diverges logarithmically upon approaching the separatrix. This divergence implies that in an ensemble of classical rotation states, each trajectory flips with a vastly different frequency even though their initial angular momenta might be very close. Thus the ensemble-averaged tennis racket flips are expected to quickly decay. 

\emph{Thermally averaged trajectory} --
As the practically most relevant scenario, we consider that the particle is initially in a displaced thermal state of mean mid-axis angular momentum $\Jo$ and temperature $T$, where $\Ay J_0^2/k_{\rm B} T \gg 1$. A lengthy and non-trivial calculation for rotational energies clearly exceeding the thermal width of the state yields that the ensemble-averaged rotor dynamics exhibit a Gaussian decay in time~\cite{supp},
\begin{subequations} \label{eq:E_Class}
\begin{eqnarray}
\left \langle \Ly(t) \right \rangle  &\simeq& J_0 \sum_{r = \pm} C_r \sum_{n = 0}^\infty \frac{(-{1})^n }{2n + 1} \cos \left [ (2n +1) \frac{2\pi t}{\tau_r} \right ] \notag \\
&& \times\exp\left[-(2n+1)^2\kappa_r^2 t^2\right] \label{eq:E_C_ass}.
\end{eqnarray}

\noindent {{where summing over $r=\pm$ covers the contributions for positive and negative separatrix energies.} The thermal period and decay rates on both sides of the separatrix are given by 
\begin{eqnarray}
\tau_{\pm} &\approx&\frac{2}{\nu_0} \ln \left (\frac{16}{3}\frac{\Ay\Jo^2}{\kb T}  \delta_{\pm} \right )  + \frac{2\gamma_{\rm EM}}{\nu_0}
,\\
\kappa_{\pm} &= & \frac{2\pi^2}{\sqrt{3} \nu_0\tau_{\pm}^2}
\end{eqnarray}

where $\nu_0 = 2\sqrt{(\Ay - \Ax)(\Az - \Ay)}\Jo$ is the characteristic frequency and $\gamma_{\rm EM} \simeq 0.58\cdots$ is the Euler-Mascheroni constant. In addition, we defined the constant
\begin{equation}
    C_+ = \frac{4}{\pi}\left[1 + \frac{3A_{3} (A_{2} - A_{1}) + \Ay (A_{3} - A_{1})}{3 A_{1} (A_{3} - A_{2}) + \Ay (A_{3} - A_{1})}\right]^{-1}
\end{equation}
and the geometry-dependent factor $\delta_{+}=1 + 3A_{3}(A_{2} - A_{1})/\Ay (A_{3}  - A_{1})$ for the trajectories above the separatrix. The values below the separatrix, $C_-$ and $\delta_-$, are obtained by exchanging $\Az$ and $\Ax$. The thermally averaged trajectory \eqref{eq:E_Class} ensures that $\langle \Ly(0)\rangle = \Jo$ and provides an accurate approximation to the exact dynamics in a wide range of asymmetric rigid rotor geometries, $10^{-2}<(\Az - \Ay)/(\Ay - \Ax)<10^2$.}
\end{subequations}

The tennis racket dynamics \eqref{eq:E_Class} are experimentally observable if the rotor is initially aligned, so that the conserved total angular momentum vector ${\bf J}$ always points in approximately the same space-fixed direction in each experimental run. Then the mid-axis angular momentum $\langle \Ly(t) \rangle = \langle {\bf J} \cdot {\bf n}_2(t)\rangle$ is revealed by measuring the mid-axis orientation ${\bf n}_2(t)$. On both sides of the separatrix the tennis racket flips decay on the timescale $1/\kappa_\pm \sim \ln^2 (A_2 J_0^2/k_{\rm B} T)/\nu_0$ and can thus only be observed if the mean rotational energy exponentially exceeds the rotational temperature. Surprisingly, this is not necessary due to the onset of quantum persistent tennis racket flipping.

\emph{Quantum tennis racket tunnelling} --
The quantum rigid body rotations of an asymmetric top are induced by the free Hamiltonian ${\sf H} = \Ax \sLx^2 + \Ay \sLy^2 + \Az \sLz^2$, with the body-fixed angular momentum components ${\sf L}_k$, satisfying $[{\sf L}_k, {\sf L}_l] = -i \hbar \varepsilon_{klm} {\sf L}_m$, and $\varepsilon_{klm}$ the Levi-Civita symbol. This Hamiltonian commutes with the total angular momentum operator ${\sf L}^2 = \sLx^2 + \sLy^2 + \sLz^2$ (with eigenvalues $\hbar^2 j(j+1)$), and thus also with the separatrix operator ${\sf S} = {\sf H} - \Ay {\sf L}^2$.

Each separatrix eigenstate $\ket{S_{jn}}$ (with $n = -j,-j+1,...,j$),  falls into one of four orthogonal subspaces, characterised by the rotational symmetry of the state \cite{landau2013quantum}. The resulting eigenvalues $S_{jn}$ are twofold degenerate for eigenstates close to the axes $\Lx$ or $\Lz$, because classical rotations around these axes are stable. However, the degeneracy is lifted close to the classically unstable mid-axis by quantum tunnelling between counter-rotating trajectories of equal energies \cite{harter1984rotational}. We will show that these tunnelling contributions are necessary for persistent quantum tennis racket flipping.  

{In order to quantify the role of tunnelling for the quantum mid-axis dynamics, it is helpful to transform the Hamiltonian to that of a linear rotor with effective moment of inertia $I_{\rm eff}$ in an effective potential. The Hamiltonian for the asymmetric rotor can be written in action angle variables, $H = (L^2 - \Lz^2) [\Ax \cos^2\varphi + \Ay \sin^2\varphi] + \Az \Lz^2$~\cite{hamraoui2018}, with $\varphi$ the conjugate coordinate to $\Lz$. The canonical point transformation via $\alpha = \pi F(\varphi,\Delta)/2 K(\Delta)$, with $\Delta = (\Ay - \Ax)/(\Az - \Ax)$, $K(\cdot)$ the complete elliptic integral of the first kind, and $F(\cdot)$ the incomplete elliptic integral of the first kind, shows that the dynamics close to the separatrix are that of a quantum planar rotor in a potential, ${\sf S}_j\simeq {\sf p}_\alpha^2/2 I_{\rm eff} + V_j^{\rm eff}( \alpha)$, for $\alpha \in [0,2\pi)$. Here, ${\sf p}_\alpha$ denotes the angular momentum operator of the planar rotor (for fixed $j$), $I_{\rm eff} = 2 \Delta K^2(\Delta) / (\Ay - \Ax)\pi^2$, and the $j$-dependent effective potential is
\begin{eqnarray}\label{eq:effpot}
 V_j^{\rm eff}(\alpha) & = & - \hbar^2 j^2 (\Ay - \Ax) {\rm cn}^2 \left [ \frac{2}{\pi} K(\Delta) \alpha , \Delta\right ],
 \label{eq:veff}
\end{eqnarray}
with cn$(\cdot)$ the Jacobi elliptic cosine.

While a classical rigid body exhibits rotational trajectories for energies above the effective potential  ($S_j >0$) and oscillatory motion for energies below the effective potential ($S_j <0$), a quantum rotor can also be reflected above the potential barrier and tunnel through it [Fig.~\ref{fig:2}(a) and (b)]. This lifts the degeneracy of energy eigenstates close to the separatrix.} The strength of the tunnelling contribution at a fixed energy $S$ is quantified by the transmission amplitude of a single through-barrier tunnelling or above-barrier reflection event,
\begin{equation}
P_j(S) = \exp \left [- \frac{\sqrt{2 I_{\rm eff}}}{\hbar} \int_{\alpha_-}^{\alpha_+} d \alpha \sqrt{V^{\rm eff}_j(\alpha)+\vert S \vert} \right ],
\label{eq:Pr_S}
\end{equation}
where $\alpha_\pm$ are the classical turning points (evaluated for $-\vert S \vert$). The tunnelling probability approaches unity close to the separatrix $S = 0$, while quickly decaying for non-zero energies [see Fig.~\ref{fig:2}(a)].

\emph{Quantum persistent flipping} --
For fair comparison with the classical treatment, we consider the dynamics of a displaced thermal state at temperature $T$ initially rotating around its mid-axis with mean angular momentum $J_0$,
\begin{equation} \label{eq:inistate}
\rho_0 = \frac{1}{Z} \exp \left ( - \frac{{\sf H}+{\sf V}_{\rm ext}}{k_{\rm B} T}+ \frac{2 \Ay J_0}{k_{\rm B} T} \sLy  \right ).
\end{equation}
{The external potential $V_{\rm ext}$ is only present initially and serves to align the body-fixed mid-axis with a well-defined space fixed direction so that the tennis-racket dynamics can be observed by measuring the rotor orientation (see above). Here, $Z$ is the partition function.}
{Fig.~\ref{fig:2}(b) plots the angular momentum sphere for fixed $j$ showing the separatrix (black line) between classical rotational (green lines) and oscillatory (pink lines) trajectories. The Wigner function~\cite{rundle2019general} of a displaced, low temperature thermal state is plotted on the sphere, taking $j = 100$, with its time evolution --} showing periodic diffusion and recombination following the separatrix {-- shown in Fig.~\ref{fig:2}(c).}

Tunnelling and above-barrier reflection become relevant for the rotor dynamics if the initial state \eqref{eq:inistate} has strong support in the region of significant tunnelling  probability \eqref{eq:Pr_S}. Using the quarter mean energy, corresponding to approximately one quarter of all states, a straightforward calculation yields the requirement $P_{J_0}(k_{\rm B} T/4) \simeq \exp (- \pi k_{\rm B} T/4\hbar \nu_0) \gtrsim 1 \%$, or equivalently $\hbar \nu_0/k_{\rm B} T \gtrsim 0.1$. This requirement depends only on the rotor aspect ratio and angular frequency, and is thus {\it independent} of its physical size or mass. This is in marked contrast to the logarithmic criterion for observing classical tennis racket flipping discussed above, which is much more restrictive for nanoscopic rotors. We will demonstrate next that strong tunnelling coincides with the occurrence of quantum persistent tennis racket flips.

Describing nanoparticle rotations close to the separatrix for realistic parameters is numerically challenging as semiclassical methods fail {in the deep tunnelling regime ~\cite{harter1984rotational,supp}}, even though macroscopically many states are occupied. Exactly computing the expectation value $\langle {\sf L}_2(t) \rangle$ requires i   ndependently diagonalising the separatrix operator of size $(2j+1)^2$ for each thermally occupied $j$-value. For nano-scale rotors at GHz rotation frequencies, $J_0/\hbar$ is on the order of tens of millions up to even billions, and since the thermal width includes macroscopically many different $j$ values, this quickly becomes numerically intractable. However, it suffices to determine the eigenvalues for the central $j = \Jo/\hbar$ and interpolate to find the possible quantum frequencies $2 \pi/\tau_{\rm qu}(S)$ as a function of $S$ \cite{supp}.

Close to the separatrix, where tunnelling becomes relevant, the possible quantum frequencies $2 \pi/\tau_{\rm qu}(S) \simeq (S_{jn} - S_{jn-2})/\hbar$ split into two branches {[blue and gray lines, Fig.~\ref{fig:2}(d)]}, from which only the lower {(blue)} is relevant (at low temperature/fast rotation rates) due to the symmetry of the initial state~\cite{supp,harter1984rotational}. Importantly, $\tau_{\rm qu}(S)$ does not diverge at the separatrix, in contrast to the classical case (orange), {because the tunnelling splitting lifts the degeneracy of the energy levels \cite{supp}}. The resulting quantum mid-axis angular momentum dynamics can be approximated by
\begin{equation} \label{eq:quantLapprx}
\langle \sLy(t) \rangle \simeq J_0 \int dS \lambda(S) \cos   \left [ \frac{2 \pi t}{\tau_{\rm qu}(S)}\right ],
\end{equation}
where {the separatrix weight} $\lambda(S)$ is well approximated by the classical probability distribution for the separatrix energy $S$ \cite{supp},
\begin{eqnarray} \label{eq:approxlambda}
\lambda(S) 
& \propto & \exp \left [\frac{S}{2}\left(\frac{\Ax}{(\Ay-\Ax)\chi_1^2}-\frac{\Az}{(\Az-\Ay)\chi_3^2}\right)\right] \notag \\
&& \times K_0\left [\frac{|S|}{2}\left(\frac{\Ax}{(\Ay-\Ax)\chi_1^2}+\frac{\Az}{(\Az-\Ay)\chi_3^2}\right)\right]
\end{eqnarray}
with $\chi_{1/3}^2=k_{\rm B}T+\hbar \Jo A_{1/3}$, which includes contributions from both the thermal width and quantum uncertainty, {and $K_0[\cdot]$ is the zeroth order modified Bessel function of the second kind.}

Fig.~\ref{fig:2}(e) shows the exact quantum simulation of the tennis racket dynamics for $\Jo/\hbar = 10^6$, while (f) shows the approximate quantum tennis racket flipping \eqref{eq:quantLapprx} for $\Jo/\hbar = 10^8$, {at the temperature $(\Az - \Ax) \hbar \Jo/k_{\rm B} T = 1.25$, where we used the values for moments of inertia $I_1 =4.{4}\times10^3$\,amu\,$\mu$m$^2$, $I_2 = 3.5 \times 10^3$\,amu\,$\mu$m$^2$, and $I_3 =1.7\times 10^3$\,amu\,$\mu$m$^2$, close to state of the art experiments~\cite{kuhn2017full,reimann2018,ahn2018optically,rashid2018precession}}. Both plots clearly demonstrate the persistence of quantum tennis racket flips in the tunnelling regime. {The simulation for such a high value of $J_0$ in Fig.~\ref{fig:2} is only possible by using the classical approximations for the overlaps Eq.~\eqref{eq:approxlambda}, and a continuous approximation of the quantum frequency as a function of separatrix energy for the central $J_0$ value \cite{supp}.} The lack of high frequency components in (f) is because \eqref{eq:quantLapprx} only accounts for the lowest energy differences, i.e. only includes low frequency components. 

\emph{Discussion} --
Mid-axis rotations of quantum asymmetric rotors exhibit strong non-classical signatures, which can be observed with cutting-edge technology pushing nanoscale rotations towards a quantum regime~\cite{kuhn2017optically,reimann2018,ahn2018optically,rashid2018precession,korobenko2014}. For a given rotational temperature, observation of the quantum dynamics requires spinning the particle with characteristic frequency $\nu_{0}$ so that $\hbar \nu_{0}/k_{\rm B}T \gtrsim 0.1$. This relation is independent of the rotor mass but only a function of its aspect ratio. For instance, observing quantum effects at GHz rotation rates requires moderate cooling to a rotational temperature of  $100$\,mK. While this is the same temperature required to see quantum effects in a GHz linear oscillator, we note the rotation rate is given by the initial state, unlike the geometry-dependent frequency of a linear oscillator. Furthermore, for nanomechanical oscillators the resulting amplitude is femtometers, while the nanorotor superposition state extends over the extent of the particle.

State-of-the-art optical techniques enable the preparation of GHz mechanical rotations with nanoscale dielectrics \cite{reimann2018,ahn2018optically}, and beyond THz for linear molecules \cite{korobenko2014}. Rotational cooling is within reach either by cavity or feedback techniques \cite{stickler2016rotranslational,seberson2019parametric}, which have been successfully implemented for cooling the centre-of-mass motion of dielectrics \cite{delic2020cooling,tebbenjohanns2020}. Three-dimensional orientation of the nanoparticles and molecules can in principle be achieved by exploiting a permanent electric or magnetic dipole moment, which aligns with an external static field \cite{rider2016search}. Angularly accelerating trapped and aligned particles to high rotation speeds has been implemented by both the radiation torque exerted by a circularly polarised laser beam \cite{kuhn2017full,reimann2018,ahn2018optically}, and by adiabatic alignment in the field of an optical centrifuge \cite{karczmarek1999optical,korobenko2014}.

Once released, the rotor starts dispersing and recurring as shown in Fig.~\ref{fig:2}. {Its alignment as a function of time can be optically measured with high precision via light scattering as demonstrated experimentally, e.g.~\cite{kuhn2017optically,hoang2016torsional,kuhn2017full,reimann2018,ahn2018optically,ahn2020ultrasensitive,rashid2018precession,supp}.} Orientational decoherence  due to scattering of gas atoms \cite{stickler2016spatio} requires an experiment with nanorotors to be performed in high vacuum, so that on average less than one collision occurs during the tennis racket flips.

In conclusion, we provided a full theoretical treatment of the classical and quantum tennis racket effect. The observation of persistent tennis racket flips will probe the quantum nature of rotations in an unparalleled regime of macroscopic moments of inertia, where the objects rotate with billions of angular momentum quanta \cite{kuhn2017full,kuhn2017optically,reimann2018,ahn2018optically,rashid2018precession}.

\emph{Acknowledgements} --
MSK and KEK acknowledge the KIST Open Research Programme and the Royal Society. 
YM is supported by the EPSRC Centre for Doctoral Training on Controlled Quantum Dynamics at Imperial College London (EP/L016524/1) and funded by the Imperial College President's PhD Scholarship. BAS acknowledges funding from the European Union’s Horizon 2020 research and innovation programme under the Marie Skłodowska-Curie grant agreement No. 841040. This work is supported by EPSRC (EP/R044082/1). 

\bibliographystyle{myapsrev}

\newpage 
\mbox{}
\newpage
\begin{widetext}
\appendix

\section{Thermally-averaged classical trajectory} The classical mid-axis trajectory is given by a Jacobi function, which close to the separatrix ($S/\Ay L^2 \ll 1$) can be approximated by a square wave 
\begin{equation} \label{eq:sw}
\Ly(t) \simeq \frac{4\Ly}{\pi} \sum_{k = 0}^\infty \frac{(-1)^k}{2 k + 1} \cos \left [(2k+1) \frac{2 \pi t}{\tau_{\rm cl}} \right ].
\end{equation}
Here, $\tau_{\rm cl}$ depends on the transverse angular momenta $\Lx$ and $\Lz$ through Eq.~(1) in the main text, and the approximation neglects any small phase offsets.

The ensemble-averaged mid-axis trajectory is calculated by integrating \eqref{eq:sw} over its initial conditions distributed according to a displaced Gibbs state, which for $\Ay J_0^2 / k_{\rm B} T \gg 1$ can be approximated as
\begin{eqnarray}
f_0(\Lx,\Ly,\Lz)  &\simeq & \frac{\sqrt{\Ax \Az}}{2\pi \kb T} \delta ( \Ly - J_0)  \exp\left[-\frac{\Ax\Lx^2 + \Az \Lz^2}{\kb T}\right].
\end{eqnarray}

To carry out the integration, we first distinguish between trajectories above ($S>0$) and below  ($S<0$) the seperatrix. Taking the $S>0$ contribution (subscript +) restricts the integration volume to $(\Ay-\Ax)\Lx^2  < (\Az-\Ay) \Lz^2$, and noting the integrand is quadratic in $\Lx$ and $\Lz$ yields
\begin{eqnarray}
\langle\Ly(t)\rangle_+ =  \frac{8 \Jo}{\pi}\sum_{k=0}^\infty  \frac{(-1)^k}{2k + 1}\frac{\sqrt{\Ax\Az}}{\pi\kb T}  \mathcal{I}_k(t)\label{eq:ensemble}
\end{eqnarray}
where 
\begin{eqnarray}
\mathcal{I}_k(t) &=&  \int_0^\infty d\Lz \int_0^{\sqrt{\frac{\Az-\Ay}{\Ay-\Ax}}\Lz } d\Lx
e^{- (\Ax\Lx^2 + \Az \Lz^2)/\kb T}\cos\left[\frac{(2 k+1)\pi\nu_0 t}{\ln[(\Az-\Ax)S/4\nu_0^2]}\right]. 
\end{eqnarray}
Note that approximating the period via the logarithm (1) is well justified in the limit of high rotation energies, $\Ay \Jo^2/\kb T  \gg 1$, where all trajectories are close to the separatrix.

The $\Lx$ integral can be approximated by $\int_0^a dxf(x) \approx a f(a/2)$. This works both for $\sqrt{(\Az-\Ay)/(\Ay-\Ax)} > 1$, where the $\Lx$ Gaussian vanishes quickly, and for $\sqrt{(\Az-\Ay)/(\Ay-\Ax)}\lesssim 1$, where the Gaussian is slowly varying. The integral thus becomes,
\begin{subequations}
	\begin{eqnarray}
	\mathcal{I}_k(t) &= &\sqrt{\frac{\Az-\Ay}{\Ay-\Ax}}\int_0^{\infty}d\Lz \Lz e^{-\Lz^2/2\sigma^2}\cos\left[\frac{(2k + 1)\pi \nu_0 }{\ln(\Lz^2/\gamma^2)}\right]
	\end{eqnarray}
	where we have introduced the constants 
	\begin{eqnarray}
	(2\sigma^2)^{-1} &=& \frac{\Az}{\kb T} + \frac{\Ax}{4\kb T}\frac{\Az-\Ay}{\Ay-\Ax},\\
	\gamma^{2} &=& \frac{64}{3}\frac{\Ay-\Ax}{\Az-\Ax} \Jo^2.
	\end{eqnarray}
\end{subequations}
Substituting $v = - \ln(\Lz^2/\gamma^2)$ yields
\begin{eqnarray}
\mathcal{I}_k(t) &=&  \sqrt{\frac{\Az-\Ay}{\Ay-\Ax}} \frac{\gamma^2}{2}\int_0^{\infty} dv \exp \left(-v-\frac{\gamma^2}{2\sigma^2}e^{-v} \right ) \cos\left[(2k + 1) \frac{\pi \nu_0  t}{v}\right].
\end{eqnarray}

The exponential in this integral may be approximated as a Gaussian, whose mean and variance are determined by  the probability density function $p(v)= \gamma^2\exp [-v-\gamma^2e^{-v}/2\sigma^2]/2\sigma^2$. This yields
\begin{subequations}
	\begin{eqnarray}
	v_{\rm m} & = & \langle v \rangle  = \ln \left ( \frac{\gamma^2}{2 \sigma^2} \right ) + \gamma_{\rm EM}, \\
	\Sigma^2 &\equiv& \langle (v - \langle v \rangle )^2 \rangle = \frac{\pi^2}{6}
	\end{eqnarray}
\end{subequations}
where we note that the variance $\Sigma^2$ is independent of $\gamma$ and $\sigma$. 

The lower bound of the integral may be extended to minus infinity (valid for $v_{\rm m} \gtrsim 3$), and the final integral over $v$ can be computed by linearising the $\cos[ (2k + 1)\pi\nu_0 t/v]$ about the central peak in the Gaussian at $v_{\rm m}$. This yields
\begin{eqnarray}
\mathcal{I}_k(t) &=& \frac{\sigma^2}{\Sigma^2}\sqrt{\frac{\Az-\Ay}{\Ay-\Ax}} \exp\left[-(2k + 1)^2\pi^4\frac{\nu_0^2 t^2}{12 v_{\rm m}^4}\right] \cos\left[(2k +1)\pi\frac{ \nu_0 t}{v_{\rm m}}\right].
\label{eq:I_k}
\end{eqnarray}

The contribution from the other side of the separatrix $S<0$ can be calculated by exchanging $\Ax$ and $\Az$. The final expression requires renormalization to the initial condition $\langle L_2(0) \rangle = \Jo$ due to the approximations involved in the derivation. Summing the two contributions and renormalizing gives (2) in the main text.
A comparison between the analytic result (2) in the main text and a full-fledged Monte-Carlo simulation is shown in Fig.~\ref{fig:classical}. It demonstrates that the approximations yield very good agreement in the relevant parameter regime.


\begin{figure}
	\centering
	\includegraphics[width=0.35\columnwidth]{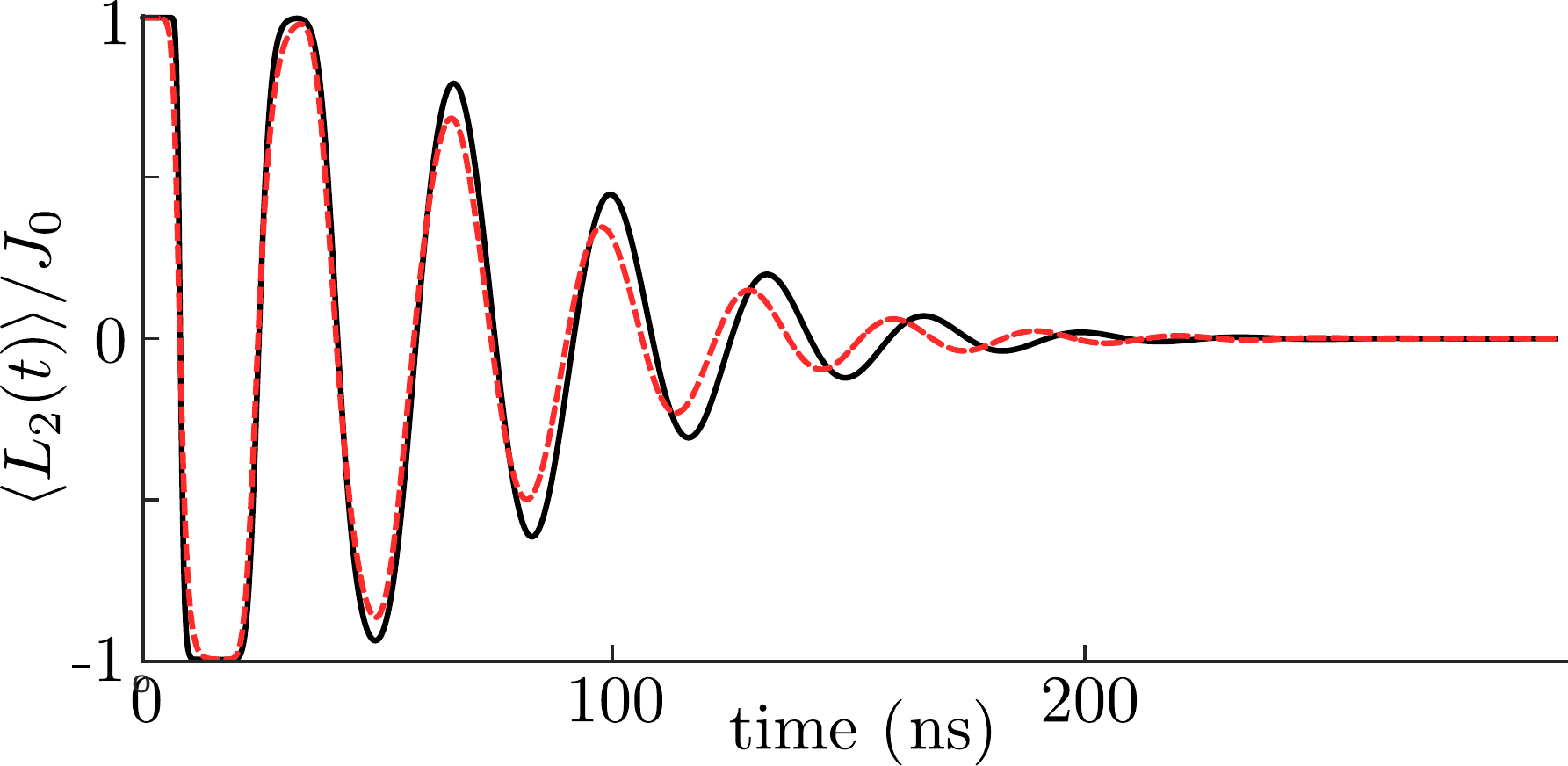}
	\caption{Comparison of (2) in the main text (solid black line) with full-fledged Monte-Carlo simulations (dashed red line) for $\Jo/\hbar = 1.2\times 10^8$ and $T =1$~K. The simulations show that Eq. (2) in describes the exact classical rotation dynamics to very high accuracy.} \label{fig:classical}
\end{figure}

\section{Eigenstate symmetries} The separatrix operator at fixed j commutes with the three $\pi$-rotation operators $R _n=\exp(i\pi L_n/\hbar)$, where $n = 1,2,3$. Thus, the separatrix eigenstates $\ket{S_k}, k = 1,... 2j + 1$ are also eigenstates of $R_n$ with eigenvalues $\pm 1$ and can therefore be grouped into four orthogonal subsets according to their simultaneous $(R_1,R_2,R_3)$ eigenvalues as as $(+,+,+)$, $(+,-,-)$, $(-,-,+)$, $(-,+,-)$. These four symmetry classes are conventionally labelled by $a_1, a_2, b_1$ or $b_2$~[C. Van Winter, {\it The asymmetric  rotator  in  quantum  mechanics}, Physica {\bf 20}, 274 (1954)].

When ordering the separatrix eigenstates in ascending order one finds that they belong to the symmetry groups depicted in Tab. \ref{tab:s1} (left) [C. Van Winter, {\it The asymmetric  rotator  in  quantum  mechanics}, Physica {\bf 20}, 274 (1954);  W. G. Harter and J. C. Mitchell, {\it Molecular eigensolutions: symmetry  analysis  and  fine  structure}, Int. J. Mol.  Sci. \textbf{14}, 714 (2013)].

\begin{table}[H]\label{tab:s1}
	\caption{Symmetry subgroups  of the separatrix eigenstates and of the initial state (5) for fixed $j$.}
	\centering
	\begin{tabular}{|c|c|c||c||c|c|c|}
		\hline
		\begin{tabular}[c]{@{}c@{}}Separatrix\\ eigenstate $k$\end{tabular} & \multicolumn{2}{c||}{\begin{tabular}[c]{@{}c@{}}$(R_1,R_2,R_3)$\\ eigenvalues\end{tabular}} & \hspace*{2cm}& \begin{tabular}[c]{@{}l@{}}initial-state \\ eigenstate $k$\end{tabular} & \multicolumn{2}{c|}{$R_2$ eigenvalue}\\ \hline
		& even $j$        & odd $j$ & & &  even $j$ & odd $j$       \\ \hline
		1                                                                 & $+++$          & $--+$   && 1                   & $+1$               & $-1$            \\ \hline
		2                                                                 & $+--$          & $-+-$  && 2                           & $-1$               & $+1$          \\ \hline
		3                                                                 & $-+-$          & $+--$  && 3              & $+1$               & $-1$      \\ \hline
		4                                                                 & $--+$          & $+++$  && 4              & $-1$               & $+1$                   \\ \hline
		5                                                                 & $+++$          & $--+$  && 5              & $+1$               & $-1$                   \\ \hline
		6                                                                 & $+--$          & $-+-$  && 6              & $-1$               & $+1$      \\ \hline
		7                                                                 & $-+-$          & $+--$  && 7              & $+1$               & $-1$                   \\ \hline
		8                                                                 & $--+$          & $+++$  && 8              & $-1$               & $+1$      \\ \hline
		...                                                               & ...            & ...    && ...            & ...                & ...               \\ \hline
		$2j + 1$                                                          & $+++$          & $+--$  && $2j + 1$       & $+1$               & $-1$               \\ \hline
	\end{tabular}
\end{table}

The eigenstates of the thermal state Eq. (5) are determined by the Hamiltonian $H_{\rm th} = A_1 L_1^2+A_2 (L_2-J_0)^2+A_3 L_3^2$, which does not commute with all three $R_n$, but only with $R_2$. Listing the eigenstates at fixed j together with their $R_2$-eigenvalues yields the right columns of Tab. \ref{tab:s1}.

As a result, for even and for odd $j$, the ground state of $H_{\rm th}$ only has support in the subspace spanned by the 1st, 3rd, 5th, 7th, etc. eigenstates of the separatrix operator, while the first excited state only has support in the subspace spanned by the 2nd, 4th, 6th, 8th, etc. eigenstates of the separatrix operator, and so on. Since for  a given $j$, the initial state is prepared close to its ground state, the dynamics are dominated by the odd eigenstates of the separatrix operator.

The dynamical frequencies are determined by the eigenvalue spacing of the separatrix operator. According to the above symmetry argument, only second nearest eigenvalues can contribute and thus, for the ground state of $H_{\rm th}$ only the odd energy differences $S_{2n+1}-S_{2n-1}$ contribute, while for the first excited state only the even energy differences $S_{2n+2}-S_{2n}$ matter. The odd differences correspond to the lower frequency quantum frequency curve in Fig. 2d , while the even differences correspond to the upper quantum frequency curve. Thus, mainly the lower frequency branch contributes to the quantum tennis racket dynamics.

Why the odd differences always yield the lower frequency curve can be understood from the properties of the rigid rotor spectrum, however the argument is a bit involved. We will give here a short version, focussing on the most asymmetric rotor with $1/I_2=(1/I_1+1/I_3 )/2$. In this case and for fixed $j$, there is always the same number of eigenstates above and below the separatrix ($S = 0$) and the corresponding eigenenergies are symmetric, i.e. the $k$-th and the $(2j+2-k)$-th state have opposite eigenvalues $S_{2j+2-k}=-S_k$. The total number of states is $2j+1$ and therefore always odd, implying that the central eigenvalue is always $S_{j+1}= 0$. This implies that for even (odd) $j$ this central state is populated (unpopulated) in the ground state of $H_{\rm th}$. In addition, we note that the states are degenerate far above and below the separatrix and can thus naturally be grouped into pairs.

We note the following three different facts for rigid rotor spectra:
\begin{enumerate}
	\item The energy spacing between two adjacent near-degenerate pairs of eigenvalues is larger when further away from the separatrix.
	\item The tunnelling induced splitting of a near-degenerate energy levels is smaller when further away from the separatrix.
	\item The splitting of a near-degenerate pair is always smaller than the spacing between neighbouring pairs, i.e. no level crossing occurs. 
\end{enumerate}

From these it follows that the odd differences $S_{2n+1}-S_{2n-1}$ are always smaller than the the two neighbouring even differences $S_{2n+2}-S_{2n}$ and $S_{2n}  - S_{2n-2}$, (see Fig. \ref{fig:R2}).

\begin{figure}
	\centering
	\includegraphics[width = 0.35\columnwidth]{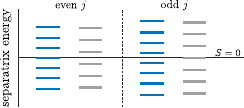} 
	\caption{Level spacing close to the separatrix $(S = 0)$. Only the blue levels are occupied in the groundstate, while the grey levels are occupied in the first excited state. For the sufficiently low temperatures $\hbar \nu_0/k_{\rm B} T \gtrsim 0.1$, the thermal state of $H_{\rm th}$ is well approximated by the convex sum of ground states of each $j$ subspace, where different $j$'s are thermally distributed. Thus only the blue states (blue curve in Fig. 2. (d) ) contribute to the dynamics.}
	\label{fig:R2}
\end{figure}

\section{Distribution of separatrix energies}

The probability density for the separatrix energy $S$ is found from the Gibbs factor by a change of variables, $\lambda(S) = \int d^3L ~\delta[S - S(\Lx,\Ly, \Lz)] \exp[-H(\Lx,\Ly,\Lz)/\kb T]/Z$, where $Z$ is the partition function. Noting $S$ only depends on $\Lx$ and $\Lz$, for a given value of $j$ we have,
\begin{eqnarray}
\lambda(S) & = & \frac{1}{N} \int d\Lx d\Lz \exp \left ( - \frac{\Ax\Lx^2}{\chi_1^2} - \frac{\Az\Lz^2}{\chi_3^2} \right ) \delta [ S - (\Az - \Ay)\Lz^2 + (\Ay - \Ax) \Lx^2],
\end{eqnarray}
where $\chi_{1/3}$ also take account of the quantum uncertainty of $L_{1/3}$. Carrying out the integral yields Eq. (7) of the main text. 

This expression can then be used to calculate the classical probability distribution of different flipping frequency components.
\begin{eqnarray}
{\rm prob}(\omega) = \int dS \delta\left( \omega - \frac{2\pi}{\tau(S)}\right)\lambda(S).
\end{eqnarray}
This probability density is shown in Fig. \ref{fig:R1} and compared with the quantum distribution, showing the tunneling-induced lifted degeneracy changes distribution of frequencies. This highlights the fact that the existence of a maximal tennis-racket flipping frequency alone is not sufficient to describe the quantum dynamics.

A comparison of the exact diagonalization to the approximations Eqs. (6) and (7) is shown in Fig. \ref{fig:R3}(b). It highlights the validity of the discussed approximations, which substantially reduces the numerical requirements. Specifically, calculating the eigenvalues and energies of the separatrix operator $S$ requires diagonalizing a $\left(2j+1\right)\times\left(2j+1\right)$-dimensional matrix for each thermally occupied $j$. The resulting $S$-eigenvectors of length $2j+1$ are required to calculate the overlap matrix elements with the initial state. The fact that the thermal width scales as $\sim~\sqrt{J_0/\hbar}$  (for $A_2\hbar J_0/k_BT={\rm const}$) makes standard numerical tools ineffective for already quite small values of $J_0$.

\begin{figure}
	\centering
	\includegraphics[width = 0.35\columnwidth]{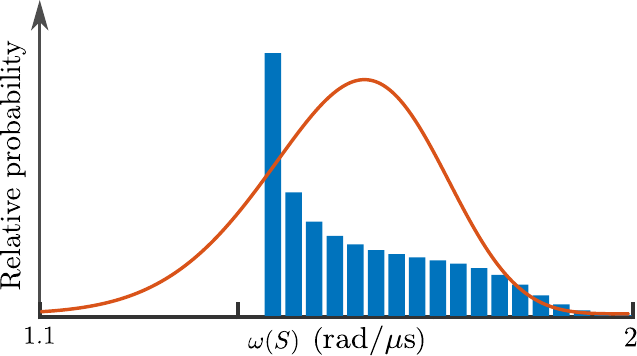}
	\caption{Classical probability density (orange curve) and quantum probability (blue bars) for flipping frequencies. The sharp peak in the quantum distribution is a consequence of the tunneling-induced lifted degeneracy and is responsible for the persistent oscillations. Parameters taken from Fig. 2 (e) : $\Jo/\hbar = 10^6$, $T = 0.07$~mK.}
	\label{fig:R1}
\end{figure}

\newpage
\section{Temperature dependence and quantum-to-classical transition}

One sees from Eq. (5) that if the rotor is close to its rotating ground state, the frequency and temperature enter mainly via the ratio $A_2\hbar J_0/k_BT$. This is also supported by the results shown in Fig. 2(e) for $~\sim 3$ MHz and $0.07$ mK and those shown in Fig. 2(f) for $\sim0.3$ GHz and 7 mK. Increasing the temperature decreases the timescale on which quantum persistent oscillations are observable, see Fig. \ref{fig:R3}(a). In the limit $A_2\hbar J_0/k_BT\ \ll 1$ the quantum dynamics approach the classical mid-axis dynamics (see Fig. \ref{fig:R5}).

One might expect semiclassical methods to become more accurate with increasing total angular momentum.  However, in the present case we consider a situation where the rotor dynamics is tightly confined to a region centred at the separatrix (classical phase space instability). In its vicinity, semiclassical methods fail even for large quantum numbers because the period of classical orbits diverges. The fact that semiclassical quantization around the separatrix produces incorrect eigenvalues for asymmetric rigid rotors has been known for a long time in the molecular physics community [W. G. Harter and J. C. Mitchell, {\it Molecular eigensolutionsymmetry  analysis  and  fine  structure}, Int. J. Mol.  Sci. \textbf{14}, 714 (2013){; A. A. Ovchinnikov, N. S. Erikhman, and K. A. Pronin. \textit{Vibrational-rotational excitations in nonlinear molecular systems}, Springer Science \& Business Media (2012)}]. Strong tunnelling contributions between classically equivalent trajectories close to the separatrix make it in practice impossible to keep track of all tunnelling paths.

It is important to note that the rotor is prepared in the deep tunnelling regime since $\Ay \hbar J_0 / \kb T \gtrsim 0.1$. Specifically, the rotor energy level spacing is approximately given by $\Ay \hbar J_0$, so that for each $j$ only a few (e.g. $\sim$ 100 for $\Jo /\hbar = 10^6$) rotation states contribute coherently to the dynamics. (However, $j$ is very large and the initial state includes many different $j$-states.)

\begin{figure}
	\centering
	\includegraphics[width=0.35\columnwidth]{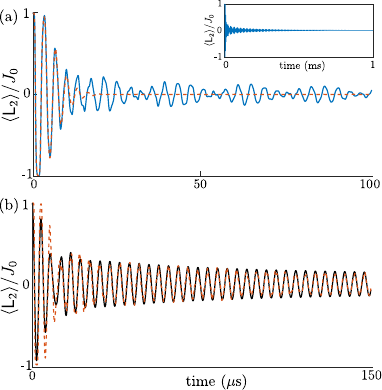}
	\caption{(a) Quantum persistent oscillations for $\Jo/\hbar = 10^6$, ($\nu_0 \approx 3$~MHz) and $T = 0.7$~mK [factor of ten higher than Fig. 2(e)]. (b) Comparison between exact diagonalization (solid black curve) and the approximations (red dashed curve) in Eqs (6) and (7). The flipping dynamics can be well described by the classical separatrix weights Eq. (7), and a continuum of frequencies Eq. (6), demonstrating that the relevant physical difference between quantum and classical tennis racket dynamics is the density of states close to the separatrix. Parameters taken from Fig. 2 (e): $\Jo/\hbar = 10^6$, $T = 0.07$~mK.}
	\label{fig:R3}
\end{figure}

\begin{figure}
	\centering
	\includegraphics[width=.35\columnwidth]{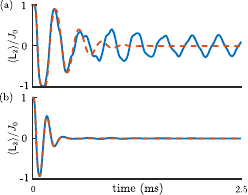}
	\caption{Transition from quantum to classical flipping dynamics for $\Jo/\hbar = 10^4$: (a) $T = 0.7$~$\mu$K. (b) $T = 70$~$\mu$K. The dependence of the flipping frequency on temperature is described by Eq. 2(b).}
	\label{fig:R5}    
\end{figure}

\section{Measuring the flipping dynamics}

The flipping motion manifests itself in the time-dependent mid-axis orientation $\mathbf{n}_2(t)$ of the rotor, as derived in the last paragraph of the section {\it Classical dynamics}. The alignment, i.e. the orientation up to a sign, of nanoscale dielectrics can be measured optically with high precision. This was demonstrated experimentally e.g. in Refs. [5,10-15,17,18]. For instance, the recent work [S.  Kuhn  et al., {\it Cavity-assisted manipulation of freely rotating silicon nanorods in high vacuum}, Nano Lett. {\bf 15}, 5604 (2015)] demonstrates excellent agreement between the detected signal and the theoretically expected orientation-dependence of the scattering light $P_{sc}\propto\sum_{k}{\chi_k^2 \left\langle{ \left[\mathbf{n}_k\left(t\right)\cdot\mathbf{e}\right]}^2\right\rangle}$ where $\bf e$ is the polarisation direction of the light beam, $\mathbf{n}_k\left(t\right)$ are the directions of the rotor principal axes, and  $\chi_k$ are the eigenvalues of the susceptibility tensor [e.g. C. F. Bohren and D. R. Huffman, Absorption and scattering of light by small particles, Wiley 2004]. This implies that direct detection of scattered light can be used to observe the second rather than the first moment of the body-fixed angular momentum with very high precision. Fig. \ref{fig:R6} demonstrates that these second moments show the same persistent flips as the first moments discussed in the manuscript. Thus, direct detection of the light scattered of a rotating nanoparticle is enough to reveal its quantum tennis racket dynamics.

\begin{figure}
	\centering
	\includegraphics[width=0.35\textwidth]{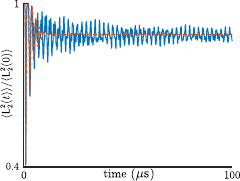}
	\caption{Comparison of second order moments, $\langle \Ly^2(t) \rangle$, for quantum (blue curve) and classical (red dashed curve) flipping dynamics using the same parameters as Fig. 2 (e) in the main text.}
	\label{fig:R6}
\end{figure}

We mention for completeness that the first moments $\left\langle L_2\left(t\right)\right\rangle \propto\langle\mathbf{n}_2\left(t\right)\cdot\mathbf{e}\rangle$ can in principle also be directly measured. Two promising routs are to  either (i) to interfere the scattered light with a bright source to realise homodyne detection or (ii) to break the inversion symmetry of the particle’s optical response along the $\mathbf{n}_2\left(t\right)$-axis, e.g. by covering one side with an optically thick layer or by attaching a marker molecule. In addition, the quantum tennis racket effect will also be relevant for small molecules. One could even expect that it will be first observed and exploited for these because large ensembles of room temperature THz superrotors are state of the art. In addition to light scattering (i.e. measuring the refractive index), there exist additional techniques to observe the orientation of small molecules such as Coulomb explosion and spectroscopy.

\end{widetext}

\begin{thebibliography}{39}%
	\makeatletter
	\providecommand \@ifxundefined [1]{%
		\@ifx{#1\undefined}
	}%
	\providecommand \@ifnum [1]{%
		\ifnum #1\expandafter \@firstoftwo
		\else \expandafter \@secondoftwo
		\fi
	}%
	\providecommand \@ifx [1]{%
		\ifx #1\expandafter \@firstoftwo
		\else \expandafter \@secondoftwo
		\fi
	}%
	\providecommand \natexlab [1]{#1}%
	\providecommand \enquote  [1]{``#1''}%
	\providecommand \bibnamefont  [1]{#1}%
	\providecommand \bibfnamefont [1]{#1}%
	\providecommand \citenamefont [1]{#1}%
	\providecommand \href@noop [0]{\@secondoftwo}%
	\providecommand \href [0]{\begingroup \@sanitize@url \@href}%
	\providecommand \@href[1]{\@@startlink{#1}\@@href}%
	\providecommand \@@href[1]{\endgroup#1\@@endlink}%
	\providecommand \@sanitize@url [0]{\catcode `\\12\catcode `\$12\catcode
		`\&12\catcode `\#12\catcode `\^12\catcode `\_12\catcode `\%12\relax}%
	\providecommand \@@startlink[1]{}%
	\providecommand \@@endlink[0]{}%
	\providecommand \url  [0]{\begingroup\@sanitize@url \@url }%
	\providecommand \@url [1]{\endgroup\@href {#1}{\urlprefix }}%
	\providecommand \urlprefix  [0]{URL }%
	\providecommand \Eprint [0]{\href }%
	\providecommand \doibase [0]{http://dx.doi.org/}%
	\providecommand \selectlanguage [0]{\@gobble}%
	\providecommand \bibinfo  [0]{\@secondoftwo}%
	\providecommand \bibfield  [0]{\@secondoftwo}%
	\providecommand \translation [1]{[#1]}%
	\providecommand \BibitemOpen [0]{}%
	\providecommand \bibitemStop [0]{}%
	\providecommand \bibitemNoStop [0]{.\EOS\space}%
	\providecommand \EOS [0]{\spacefactor3000\relax}%
	\providecommand \BibitemShut  [1]{\csname bibitem#1\endcsname}%
	\let\auto@bib@innerbib\@empty
	\bibitem [{\citenamefont {Aspelmeyer}\ \emph {et~al.}(2014)\citenamefont
		{Aspelmeyer}, \citenamefont {Kippenberg},\ and\ \citenamefont
		{Marquardt}}]{aspelmeyer2014cavity}%
	\BibitemOpen
	\bibfield  {author} {\bibinfo {author} {\bibfnamefont {Markus}\ \bibnamefont
			{Aspelmeyer}}, \bibinfo {author} {\bibfnamefont {Tobias~J}\ \bibnamefont
			{Kippenberg}}, \ and\ \bibinfo {author} {\bibfnamefont {Florian}\
			\bibnamefont {Marquardt}},\ }\bibfield  {title} {\enquote {\bibinfo {title}
			{Cavity optomechanics},}\ }\href@noop {} {\bibfield  {journal} {\bibinfo
			{journal} {Rev. Mod. Phys.}\ }\textbf {\bibinfo {volume} {86}},\ \bibinfo
		{pages} {1391} (\bibinfo {year} {2014})}\BibitemShut {NoStop}%
	\bibitem [{\citenamefont {Millen}\ \emph {et~al.}(2019)\citenamefont {Millen},
		\citenamefont {Monteiro}, \citenamefont {Pettit},\ and\ \citenamefont
		{Vamivakas}}]{millen2019optomechanics}%
	\BibitemOpen
	\bibfield  {author} {\bibinfo {author} {\bibfnamefont {James}\ \bibnamefont
			{Millen}}, \bibinfo {author} {\bibfnamefont {Tania~S}\ \bibnamefont
			{Monteiro}}, \bibinfo {author} {\bibfnamefont {Robert}\ \bibnamefont
			{Pettit}}, \ and\ \bibinfo {author} {\bibfnamefont {A~Nick}\ \bibnamefont
			{Vamivakas}},\ }\bibfield  {title} {\enquote {\bibinfo {title} {Optomechanics
				with levitated particles},}\ }\href@noop {} {\bibfield  {journal} {\bibinfo
			{journal} {arXiv preprint arXiv:1907.08198}\ } (\bibinfo {year}
		{2019})}\BibitemShut {NoStop}%
	\bibitem [{\citenamefont {Ranjit}\ \emph {et~al.}(2016)\citenamefont {Ranjit},
		\citenamefont {Cunningham}, \citenamefont {Casey},\ and\ \citenamefont
		{Geraci}}]{ranjit2016zeptonewton}%
	\BibitemOpen
	\bibfield  {author} {\bibinfo {author} {\bibfnamefont {Gambhir}\ \bibnamefont
			{Ranjit}}, \bibinfo {author} {\bibfnamefont {Mark}\ \bibnamefont
			{Cunningham}}, \bibinfo {author} {\bibfnamefont {Kirsten}\ \bibnamefont
			{Casey}}, \ and\ \bibinfo {author} {\bibfnamefont {Andrew~A}\ \bibnamefont
			{Geraci}},\ }\bibfield  {title} {\enquote {\bibinfo {title} {Zeptonewton
				force sensing with nanospheres in an optical lattice},}\ }\href@noop {}
	{\bibfield  {journal} {\bibinfo  {journal} {Phys. Rev. A}\ }\textbf {\bibinfo
			{volume} {93}},\ \bibinfo {pages} {053801} (\bibinfo {year}
		{2016})}\BibitemShut {NoStop}%
	\bibitem [{\citenamefont {Chaste}\ \emph {et~al.}(2012)\citenamefont {Chaste},
		\citenamefont {Eichler}, \citenamefont {Moser}, \citenamefont {Ceballos},
		\citenamefont {Rurali},\ and\ \citenamefont
		{Bachtold}}]{chaste2012nanomechanical}%
	\BibitemOpen
	\bibfield  {author} {\bibinfo {author} {\bibfnamefont {Julien}\ \bibnamefont
			{Chaste}}, \bibinfo {author} {\bibfnamefont {A}~\bibnamefont {Eichler}},
		\bibinfo {author} {\bibfnamefont {J}~\bibnamefont {Moser}}, \bibinfo {author}
		{\bibfnamefont {G}~\bibnamefont {Ceballos}}, \bibinfo {author} {\bibfnamefont
			{R}~\bibnamefont {Rurali}}, \ and\ \bibinfo {author} {\bibfnamefont
			{A}~\bibnamefont {Bachtold}},\ }\bibfield  {title} {\enquote {\bibinfo
			{title} {A nanomechanical mass sensor with yoctogram resolution},}\
	}\href@noop {} {\bibfield  {journal} {\bibinfo  {journal} {Nat. Nanotechn.}\
		}\textbf {\bibinfo {volume} {7}},\ \bibinfo {pages} {301} (\bibinfo {year}
		{2012})}\BibitemShut {NoStop}%
	\bibitem [{\citenamefont {Kuhn}\ \emph
		{et~al.}(2017{\natexlab{a}})\citenamefont {Kuhn}, \citenamefont {Stickler},
		\citenamefont {Kosloff}, \citenamefont {Patolsky}, \citenamefont
		{Hornberger}, \citenamefont {Arndt},\ and\ \citenamefont
		{Millen}}]{kuhn2017optically}%
	\BibitemOpen
	\bibfield  {author} {\bibinfo {author} {\bibfnamefont {Stefan}\ \bibnamefont
			{Kuhn}}, \bibinfo {author} {\bibfnamefont {Benjamin~A}\ \bibnamefont
			{Stickler}}, \bibinfo {author} {\bibfnamefont {Alon}\ \bibnamefont
			{Kosloff}}, \bibinfo {author} {\bibfnamefont {Fernando}\ \bibnamefont
			{Patolsky}}, \bibinfo {author} {\bibfnamefont {Klaus}\ \bibnamefont
			{Hornberger}}, \bibinfo {author} {\bibfnamefont {Markus}\ \bibnamefont
			{Arndt}}, \ and\ \bibinfo {author} {\bibfnamefont {James}\ \bibnamefont
			{Millen}},\ }\bibfield  {title} {\enquote {\bibinfo {title} {Optically driven
				ultra-stable nanomechanical rotor},}\ }\href@noop {} {\bibfield  {journal}
		{\bibinfo  {journal} {Nat. Commun.}\ }\textbf {\bibinfo {volume} {8}},\
		\bibinfo {pages} {1670} (\bibinfo {year} {2017}{\natexlab{a}})}\BibitemShut
	{NoStop}%
	\bibitem [{\citenamefont {Romero-Isart}\ \emph {et~al.}(2011)\citenamefont
		{Romero-Isart}, \citenamefont {Pflanzer}, \citenamefont {Blaser},
		\citenamefont {Kaltenbaek}, \citenamefont {Kiesel}, \citenamefont
		{Aspelmeyer},\ and\ \citenamefont {Cirac}}]{romero2011large}%
	\BibitemOpen
	\bibfield  {author} {\bibinfo {author} {\bibfnamefont {Oriol}\ \bibnamefont
			{Romero-Isart}}, \bibinfo {author} {\bibfnamefont {Anika~C}\ \bibnamefont
			{Pflanzer}}, \bibinfo {author} {\bibfnamefont {Florian}\ \bibnamefont
			{Blaser}}, \bibinfo {author} {\bibfnamefont {Rainer}\ \bibnamefont
			{Kaltenbaek}}, \bibinfo {author} {\bibfnamefont {Nikolai}\ \bibnamefont
			{Kiesel}}, \bibinfo {author} {\bibfnamefont {Markus}\ \bibnamefont
			{Aspelmeyer}}, \ and\ \bibinfo {author} {\bibfnamefont {J~Ignacio}\
			\bibnamefont {Cirac}},\ }\bibfield  {title} {\enquote {\bibinfo {title}
			{Large quantum superpositions and interference of massive nanometer-sized
				objects},}\ }\href@noop {} {\bibfield  {journal} {\bibinfo  {journal} {Phys.
				Rev. Lett.}\ }\textbf {\bibinfo {volume} {107}},\ \bibinfo {pages} {020405}
		(\bibinfo {year} {2011})}\BibitemShut {NoStop}%
	\bibitem [{\citenamefont {Scala}\ \emph {et~al.}(2013)\citenamefont {Scala},
		\citenamefont {Kim}, \citenamefont {Morley}, \citenamefont {Barker},\ and\
		\citenamefont {Bose}}]{scala2013matter}%
	\BibitemOpen
	\bibfield  {author} {\bibinfo {author} {\bibfnamefont {M.}~\bibnamefont
			{Scala}}, \bibinfo {author} {\bibfnamefont {M.~S.}\ \bibnamefont {Kim}},
		\bibinfo {author} {\bibfnamefont {G.~W.}\ \bibnamefont {Morley}}, \bibinfo
		{author} {\bibfnamefont {P.~F.}\ \bibnamefont {Barker}}, \ and\ \bibinfo
		{author} {\bibfnamefont {S.}~\bibnamefont {Bose}},\ }\bibfield  {title}
	{\enquote {\bibinfo {title} {Matter-wave interferometry of a levitated
				thermal nano-oscillator induced and probed by a spin},}\ }\href@noop {}
	{\bibfield  {journal} {\bibinfo  {journal} {Phys. Rev. Lett.}\ }\textbf
		{\bibinfo {volume} {111}},\ \bibinfo {pages} {180403} (\bibinfo {year}
		{2013})}\BibitemShut {NoStop}%
	\bibitem [{\citenamefont {Bateman}\ \emph {et~al.}(2014)\citenamefont
		{Bateman}, \citenamefont {Nimmrichter}, \citenamefont {Hornberger},\ and\
		\citenamefont {Ulbricht}}]{bateman2014near}%
	\BibitemOpen
	\bibfield  {author} {\bibinfo {author} {\bibfnamefont {James}\ \bibnamefont
			{Bateman}}, \bibinfo {author} {\bibfnamefont {Stefan}\ \bibnamefont
			{Nimmrichter}}, \bibinfo {author} {\bibfnamefont {Klaus}\ \bibnamefont
			{Hornberger}}, \ and\ \bibinfo {author} {\bibfnamefont {Hendrik}\
			\bibnamefont {Ulbricht}},\ }\bibfield  {title} {\enquote {\bibinfo {title}
			{Near-field interferometry of a free-falling nanoparticle from a point-like
				source},}\ }\href@noop {} {\bibfield  {journal} {\bibinfo  {journal} {Nat.
				Commun.}\ }\textbf {\bibinfo {volume} {5}},\ \bibinfo {pages} {4788}
		(\bibinfo {year} {2014})}\BibitemShut {NoStop}%
	\bibitem [{\citenamefont {Stickler}\ \emph {et~al.}(2018)\citenamefont
		{Stickler}, \citenamefont {Papendell}, \citenamefont {Kuhn}, \citenamefont
		{Schrinski}, \citenamefont {Millen}, \citenamefont {Arndt},\ and\
		\citenamefont {Hornberger}}]{stickler2018probing}%
	\BibitemOpen
	\bibfield  {author} {\bibinfo {author} {\bibfnamefont {Benjamin~A}\
			\bibnamefont {Stickler}}, \bibinfo {author} {\bibfnamefont {Birthe}\
			\bibnamefont {Papendell}}, \bibinfo {author} {\bibfnamefont {Stefan}\
			\bibnamefont {Kuhn}}, \bibinfo {author} {\bibfnamefont {Bj{\"o}rn}\
			\bibnamefont {Schrinski}}, \bibinfo {author} {\bibfnamefont {James}\
			\bibnamefont {Millen}}, \bibinfo {author} {\bibfnamefont {Markus}\
			\bibnamefont {Arndt}}, \ and\ \bibinfo {author} {\bibfnamefont {Klaus}\
			\bibnamefont {Hornberger}},\ }\bibfield  {title} {\enquote {\bibinfo {title}
			{Probing macroscopic quantum superpositions with nanorotors},}\ }\href@noop
	{} {\bibfield  {journal} {\bibinfo  {journal} {New J. Phys.}\ }\textbf
		{\bibinfo {volume} {20}},\ \bibinfo {pages} {122001} (\bibinfo {year}
		{2018})}\BibitemShut {NoStop}%
	\bibitem [{\citenamefont {Hoang}\ \emph {et~al.}(2016)\citenamefont {Hoang},
		\citenamefont {Ma}, \citenamefont {Ahn}, \citenamefont {Bang}, \citenamefont
		{Robicheaux}, \citenamefont {Yin},\ and\ \citenamefont
		{Li}}]{hoang2016torsional}%
	\BibitemOpen
	\bibfield  {author} {\bibinfo {author} {\bibfnamefont {Thai~M}\ \bibnamefont
			{Hoang}}, \bibinfo {author} {\bibfnamefont {Yue}\ \bibnamefont {Ma}},
		\bibinfo {author} {\bibfnamefont {Jonghoon}\ \bibnamefont {Ahn}}, \bibinfo
		{author} {\bibfnamefont {Jaehoon}\ \bibnamefont {Bang}}, \bibinfo {author}
		{\bibfnamefont {F}~\bibnamefont {Robicheaux}}, \bibinfo {author}
		{\bibfnamefont {Zhang-Qi}\ \bibnamefont {Yin}}, \ and\ \bibinfo {author}
		{\bibfnamefont {Tongcang}\ \bibnamefont {Li}},\ }\bibfield  {title} {\enquote
		{\bibinfo {title} {Torsional optomechanics of a levitated nonspherical
				nanoparticle},}\ }\href@noop {} {\bibfield  {journal} {\bibinfo  {journal}
			{Phys. Rev. Lett.}\ }\textbf {\bibinfo {volume} {117}},\ \bibinfo {pages}
		{123604} (\bibinfo {year} {2016})}\BibitemShut {NoStop}%
	\bibitem [{\citenamefont {Kuhn}\ \emph
		{et~al.}(2017{\natexlab{b}})\citenamefont {Kuhn}, \citenamefont {Kosloff},
		\citenamefont {Stickler}, \citenamefont {Patolsky}, \citenamefont
		{Hornberger}, \citenamefont {Arndt},\ and\ \citenamefont
		{Millen}}]{kuhn2017full}%
	\BibitemOpen
	\bibfield  {author} {\bibinfo {author} {\bibfnamefont {Stefan}\ \bibnamefont
			{Kuhn}}, \bibinfo {author} {\bibfnamefont {Alon}\ \bibnamefont {Kosloff}},
		\bibinfo {author} {\bibfnamefont {Benjamin~A}\ \bibnamefont {Stickler}},
		\bibinfo {author} {\bibfnamefont {Fernando}\ \bibnamefont {Patolsky}},
		\bibinfo {author} {\bibfnamefont {Klaus}\ \bibnamefont {Hornberger}},
		\bibinfo {author} {\bibfnamefont {Markus}\ \bibnamefont {Arndt}}, \ and\
		\bibinfo {author} {\bibfnamefont {James}\ \bibnamefont {Millen}},\ }\bibfield
	{title} {\enquote {\bibinfo {title} {Full rotational control of levitated
				silicon nanorods},}\ }\href@noop {} {\bibfield  {journal} {\bibinfo
			{journal} {Optica}\ }\textbf {\bibinfo {volume} {4}},\ \bibinfo {pages}
		{356--360} (\bibinfo {year} {2017}{\natexlab{b}})}\BibitemShut {NoStop}%
	\bibitem [{\citenamefont {Reimann}\ \emph {et~al.}(2018)\citenamefont
		{Reimann}, \citenamefont {Doderer}, \citenamefont {Hebestreit}, \citenamefont
		{Diehl}, \citenamefont {Frimmer}, \citenamefont {Windey}, \citenamefont
		{Tebbenjohanns},\ and\ \citenamefont {Novotny}}]{reimann2018}%
	\BibitemOpen
	\bibfield  {author} {\bibinfo {author} {\bibfnamefont {Ren\'e}\ \bibnamefont
			{Reimann}}, \bibinfo {author} {\bibfnamefont {Michael}\ \bibnamefont
			{Doderer}}, \bibinfo {author} {\bibfnamefont {Erik}\ \bibnamefont
			{Hebestreit}}, \bibinfo {author} {\bibfnamefont {Rozenn}\ \bibnamefont
			{Diehl}}, \bibinfo {author} {\bibfnamefont {Martin}\ \bibnamefont {Frimmer}},
		\bibinfo {author} {\bibfnamefont {Dominik}\ \bibnamefont {Windey}}, \bibinfo
		{author} {\bibfnamefont {Felix}\ \bibnamefont {Tebbenjohanns}}, \ and\
		\bibinfo {author} {\bibfnamefont {Lukas}\ \bibnamefont {Novotny}},\
	}\bibfield  {title} {\enquote {\bibinfo {title} {Ghz rotation of an optically
				trapped nanoparticle in vacuum},}\ }\href {\doibase
		10.1103/PhysRevLett.121.033602} {\bibfield  {journal} {\bibinfo  {journal}
			{Phys. Rev. Lett.}\ }\textbf {\bibinfo {volume} {121}},\ \bibinfo {pages}
		{033602} (\bibinfo {year} {2018})}\BibitemShut {NoStop}%
	\bibitem [{\citenamefont {Ahn}\ \emph {et~al.}(2018)\citenamefont {Ahn},
		\citenamefont {Xu}, \citenamefont {Bang}, \citenamefont {Deng}, \citenamefont
		{Hoang}, \citenamefont {Han}, \citenamefont {Ma},\ and\ \citenamefont
		{Li}}]{ahn2018optically}%
	\BibitemOpen
	\bibfield  {author} {\bibinfo {author} {\bibfnamefont {Jonghoon}\
			\bibnamefont {Ahn}}, \bibinfo {author} {\bibfnamefont {Zhujing}\ \bibnamefont
			{Xu}}, \bibinfo {author} {\bibfnamefont {Jaehoon}\ \bibnamefont {Bang}},
		\bibinfo {author} {\bibfnamefont {Yu-Hao}\ \bibnamefont {Deng}}, \bibinfo
		{author} {\bibfnamefont {Thai~M}\ \bibnamefont {Hoang}}, \bibinfo {author}
		{\bibfnamefont {Qinkai}\ \bibnamefont {Han}}, \bibinfo {author}
		{\bibfnamefont {Ren-Min}\ \bibnamefont {Ma}}, \ and\ \bibinfo {author}
		{\bibfnamefont {Tongcang}\ \bibnamefont {Li}},\ }\bibfield  {title} {\enquote
		{\bibinfo {title} {Optically levitated nanodumbbell torsion balance and ghz
				nanomechanical rotor},}\ }\href@noop {} {\bibfield  {journal} {\bibinfo
			{journal} {Phys. Rev. Lett.}\ }\textbf {\bibinfo {volume} {121}},\ \bibinfo
		{pages} {033603} (\bibinfo {year} {2018})}\BibitemShut {NoStop}%
	\bibitem [{\citenamefont {Ahn}\ \emph {et~al.}(2020)\citenamefont {Ahn},
		\citenamefont {Xu}, \citenamefont {Bang}, \citenamefont {Ju}, \citenamefont
		{Gao},\ and\ \citenamefont {Li}}]{ahn2020ultrasensitive}%
	\BibitemOpen
	\bibfield  {author} {\bibinfo {author} {\bibfnamefont {Jonghoon}\
			\bibnamefont {Ahn}}, \bibinfo {author} {\bibfnamefont {Zhujing}\ \bibnamefont
			{Xu}}, \bibinfo {author} {\bibfnamefont {Jaehoon}\ \bibnamefont {Bang}},
		\bibinfo {author} {\bibfnamefont {Peng}\ \bibnamefont {Ju}}, \bibinfo
		{author} {\bibfnamefont {Xingyu}\ \bibnamefont {Gao}}, \ and\ \bibinfo
		{author} {\bibfnamefont {Tongcang}\ \bibnamefont {Li}},\ }\bibfield  {title}
	{\enquote {\bibinfo {title} {Ultrasensitive torque detection with an
				optically levitated nanorotor},}\ }\href@noop {} {\bibfield  {journal}
		{\bibinfo  {journal} {Nat. Nanotechnol.}\ ,\ \bibinfo {pages} {1--5}}
		(\bibinfo {year} {2020})}\BibitemShut {NoStop}%
	\bibitem [{\citenamefont {Rashid}\ \emph {et~al.}(2018)\citenamefont {Rashid},
		\citenamefont {Toro{\v{s}}}, \citenamefont {Setter},\ and\ \citenamefont
		{Ulbricht}}]{rashid2018precession}%
	\BibitemOpen
	\bibfield  {author} {\bibinfo {author} {\bibfnamefont {Muddassar}\
			\bibnamefont {Rashid}}, \bibinfo {author} {\bibfnamefont {Marko}\
			\bibnamefont {Toro{\v{s}}}}, \bibinfo {author} {\bibfnamefont {Ashley}\
			\bibnamefont {Setter}}, \ and\ \bibinfo {author} {\bibfnamefont {Hendrik}\
			\bibnamefont {Ulbricht}},\ }\bibfield  {title} {\enquote {\bibinfo {title}
			{Precession motion in levitated optomechanics},}\ }\href@noop {} {\bibfield
		{journal} {\bibinfo  {journal} {Phys. Rev. Lett.}\ }\textbf {\bibinfo
			{volume} {121}},\ \bibinfo {pages} {253601} (\bibinfo {year}
		{2018})}\BibitemShut {NoStop}%
	\bibitem [{\citenamefont {Deli{\'c}}\ \emph {et~al.}(2020)\citenamefont
		{Deli{\'c}}, \citenamefont {Reisenbauer}, \citenamefont {Dare}, \citenamefont
		{Grass}, \citenamefont {Vuleti{\'c}}, \citenamefont {Kiesel},\ and\
		\citenamefont {Aspelmeyer}}]{delic2020cooling}%
	\BibitemOpen
	\bibfield  {author} {\bibinfo {author} {\bibfnamefont {Uro{\v{s}}}\
			\bibnamefont {Deli{\'c}}}, \bibinfo {author} {\bibfnamefont {Manuel}\
			\bibnamefont {Reisenbauer}}, \bibinfo {author} {\bibfnamefont {Kahan}\
			\bibnamefont {Dare}}, \bibinfo {author} {\bibfnamefont {David}\ \bibnamefont
			{Grass}}, \bibinfo {author} {\bibfnamefont {Vladan}\ \bibnamefont
			{Vuleti{\'c}}}, \bibinfo {author} {\bibfnamefont {Nikolai}\ \bibnamefont
			{Kiesel}}, \ and\ \bibinfo {author} {\bibfnamefont {Markus}\ \bibnamefont
			{Aspelmeyer}},\ }\bibfield  {title} {\enquote {\bibinfo {title} {Cooling of a
				levitated nanoparticle to the motional quantum ground state},}\ }\href@noop
	{} {\bibfield  {journal} {\bibinfo  {journal} {Science}\ } (\bibinfo {year}
		{2020})}\BibitemShut {NoStop}%
	\bibitem [{\citenamefont {Delord}\ \emph {et~al.}(2020)\citenamefont {Delord},
		\citenamefont {Huillery}, \citenamefont {Nicolas},\ and\ \citenamefont
		{H{\'e}tet}}]{delord2020spin}%
	\BibitemOpen
	\bibfield  {author} {\bibinfo {author} {\bibfnamefont {T}~\bibnamefont
			{Delord}}, \bibinfo {author} {\bibfnamefont {P}~\bibnamefont {Huillery}},
		\bibinfo {author} {\bibfnamefont {L}~\bibnamefont {Nicolas}}, \ and\ \bibinfo
		{author} {\bibfnamefont {G}~\bibnamefont {H{\'e}tet}},\ }\bibfield  {title}
	{\enquote {\bibinfo {title} {Spin-cooling of the motion of a trapped
				diamond},}\ }\href@noop {} {\bibfield  {journal} {\bibinfo  {journal}
			{Nature}\ }\textbf {\bibinfo {volume} {580}},\ \bibinfo {pages} {56--59}
		(\bibinfo {year} {2020})}\BibitemShut {NoStop}%
	\bibitem [{\citenamefont {Bang}\ \emph {et~al.}(2020)\citenamefont {Bang},
		\citenamefont {Seberson}, \citenamefont {Ju}, \citenamefont {Ahn},
		\citenamefont {Xu}, \citenamefont {Gao}, \citenamefont {Robicheaux},\ and\
		\citenamefont {Li}}]{bang20205d}%
	\BibitemOpen
	\bibfield  {author} {\bibinfo {author} {\bibfnamefont {Jaehoon}\ \bibnamefont
			{Bang}}, \bibinfo {author} {\bibfnamefont {Troy}\ \bibnamefont {Seberson}},
		\bibinfo {author} {\bibfnamefont {Peng}\ \bibnamefont {Ju}}, \bibinfo
		{author} {\bibfnamefont {Jonghoon}\ \bibnamefont {Ahn}}, \bibinfo {author}
		{\bibfnamefont {Zhujing}\ \bibnamefont {Xu}}, \bibinfo {author}
		{\bibfnamefont {Xingyu}\ \bibnamefont {Gao}}, \bibinfo {author}
		{\bibfnamefont {Francis}\ \bibnamefont {Robicheaux}}, \ and\ \bibinfo
		{author} {\bibfnamefont {Tongcang}\ \bibnamefont {Li}},\ }\bibfield  {title}
	{\enquote {\bibinfo {title} {5d cooling and nonlinear dynamics of an
				optically levitated nanodumbbell},}\ }\href@noop {} {\bibfield  {journal}
		{\bibinfo  {journal} {arXiv preprint arXiv:2004.02384}\ } (\bibinfo {year}
		{2020})}\BibitemShut {NoStop}%
	\bibitem [{\citenamefont {Stickler}\ \emph
		{et~al.}(2016{\natexlab{a}})\citenamefont {Stickler}, \citenamefont
		{Nimmrichter}, \citenamefont {Martinetz}, \citenamefont {Kuhn}, \citenamefont
		{Arndt},\ and\ \citenamefont {Hornberger}}]{stickler2016rotranslational}%
	\BibitemOpen
	\bibfield  {author} {\bibinfo {author} {\bibfnamefont {Benjamin~A}\
			\bibnamefont {Stickler}}, \bibinfo {author} {\bibfnamefont {Stefan}\
			\bibnamefont {Nimmrichter}}, \bibinfo {author} {\bibfnamefont {Lukas}\
			\bibnamefont {Martinetz}}, \bibinfo {author} {\bibfnamefont {Stefan}\
			\bibnamefont {Kuhn}}, \bibinfo {author} {\bibfnamefont {Markus}\ \bibnamefont
			{Arndt}}, \ and\ \bibinfo {author} {\bibfnamefont {Klaus}\ \bibnamefont
			{Hornberger}},\ }\bibfield  {title} {\enquote {\bibinfo {title}
			{Rotranslational cavity cooling of dielectric rods and disks},}\ }\href@noop
	{} {\bibfield  {journal} {\bibinfo  {journal} {Phys. Rev. A}\ }\textbf
		{\bibinfo {volume} {94}},\ \bibinfo {pages} {033818} (\bibinfo {year}
		{2016}{\natexlab{a}})}\BibitemShut {NoStop}%
	\bibitem [{\citenamefont {Seberson}\ and\ \citenamefont
		{Robicheaux}(2019)}]{seberson2019parametric}%
	\BibitemOpen
	\bibfield  {author} {\bibinfo {author} {\bibfnamefont {T}~\bibnamefont
			{Seberson}}\ and\ \bibinfo {author} {\bibfnamefont {F}~\bibnamefont
			{Robicheaux}},\ }\bibfield  {title} {\enquote {\bibinfo {title} {Parametric
				feedback cooling of rigid body nanodumbbells in levitated optomechanics},}\
	}\href@noop {} {\bibfield  {journal} {\bibinfo  {journal} {Phys. Rev. A}\
		}\textbf {\bibinfo {volume} {99}},\ \bibinfo {pages} {013821} (\bibinfo
		{year} {2019})}\BibitemShut {NoStop}%
	\bibitem [{\citenamefont {Ma}\ \emph {et~al.}(2017)\citenamefont {Ma},
		\citenamefont {Hoang}, \citenamefont {Gong}, \citenamefont {Li},\ and\
		\citenamefont {Yin}}]{ma2017proposal}%
	\BibitemOpen
	\bibfield  {author} {\bibinfo {author} {\bibfnamefont {Yue}\ \bibnamefont
			{Ma}}, \bibinfo {author} {\bibfnamefont {Thai~M}\ \bibnamefont {Hoang}},
		\bibinfo {author} {\bibfnamefont {Ming}\ \bibnamefont {Gong}}, \bibinfo
		{author} {\bibfnamefont {Tongcang}\ \bibnamefont {Li}}, \ and\ \bibinfo
		{author} {\bibfnamefont {Zhang-qi}\ \bibnamefont {Yin}},\ }\bibfield  {title}
	{\enquote {\bibinfo {title} {Proposal for quantum many-body simulation and
				torsional matter-wave interferometry with a levitated nanodiamond},}\
	}\href@noop {} {\bibfield  {journal} {\bibinfo  {journal} {Phys. Rev. A}\
		}\textbf {\bibinfo {volume} {96}},\ \bibinfo {pages} {023827} (\bibinfo
		{year} {2017})}\BibitemShut {NoStop}%
	\bibitem [{\citenamefont {H\"ummer}\ \emph {et~al.}(2020)\citenamefont
		{H\"ummer}, \citenamefont {Lampert}, \citenamefont {Kustura}, \citenamefont
		{Maurer}, \citenamefont {Gonzalez-Ballestero},\ and\ \citenamefont
		{Romero-Isart}}]{hummer2020}%
	\BibitemOpen
	\bibfield  {author} {\bibinfo {author} {\bibfnamefont {Daniel}\ \bibnamefont
			{H\"ummer}}, \bibinfo {author} {\bibfnamefont {Ren\'e}\ \bibnamefont
			{Lampert}}, \bibinfo {author} {\bibfnamefont {Katja}\ \bibnamefont
			{Kustura}}, \bibinfo {author} {\bibfnamefont {Patrick}\ \bibnamefont
			{Maurer}}, \bibinfo {author} {\bibfnamefont {Carlos}\ \bibnamefont
			{Gonzalez-Ballestero}}, \ and\ \bibinfo {author} {\bibfnamefont {Oriol}\
			\bibnamefont {Romero-Isart}},\ }\bibfield  {title} {\enquote {\bibinfo
			{title} {Acoustic and optical properties of a fast-spinning dielectric
				nanoparticle},}\ }\href {\doibase 10.1103/PhysRevB.101.205416} {\bibfield
		{journal} {\bibinfo  {journal} {Phys. Rev. B}\ }\textbf {\bibinfo {volume}
			{101}},\ \bibinfo {pages} {205416} (\bibinfo {year} {2020})}\BibitemShut
	{NoStop}%
	\bibitem [{\citenamefont {Landau}\ and\ \citenamefont
		{Lifshitz}(1976)}]{landau1976mechanics}%
	\BibitemOpen
	\bibfield  {author} {\bibinfo {author} {\bibfnamefont {LD}~\bibnamefont
			{Landau}}\ and\ \bibinfo {author} {\bibfnamefont {EM}~\bibnamefont
			{Lifshitz}},\ }\href@noop {} {\emph {\bibinfo {title} {Course of Theoretical
				Physics 1: Mechanics}}}\ (\bibinfo  {publisher} {Pergamon Press - Oxford},\
	\bibinfo {year} {1976})\BibitemShut {NoStop}%
	\bibitem [{\citenamefont {Poinsot}(1834)}]{Poinsot}%
	\BibitemOpen
	\bibfield  {author} {\bibinfo {author} {\bibfnamefont {L}~\bibnamefont
			{Poinsot}},\ }\bibfield  {title} {\enquote {\bibinfo {title} {Th{\'e}orie
				novelle de la rotation des corps, l’institut},}\ }\href@noop {} {\bibfield
		{journal} {\bibinfo  {journal} {J. G{\'e}n{\'e}ral des Soci{\'e}t{\'e}s et
				Travaux Scientifiques}\ }\textbf {\bibinfo {volume} {2}} (\bibinfo {year}
		{1834})}\BibitemShut {NoStop}%
	\bibitem [{\citenamefont {Damme}\ \emph {et~al.}(2017)\citenamefont {Damme},
		\citenamefont {Leiner}, \citenamefont {Mardešić}, \citenamefont {Glaser},\
		and\ \citenamefont {Sugny}}]{damme_linking_2017}%
	\BibitemOpen
	\bibfield  {author} {\bibinfo {author} {\bibfnamefont {L.~Van}\ \bibnamefont
			{Damme}}, \bibinfo {author} {\bibfnamefont {D.}~\bibnamefont {Leiner}},
		\bibinfo {author} {\bibfnamefont {P.}~\bibnamefont {Mardešić}}, \bibinfo
		{author} {\bibfnamefont {S.~J.}\ \bibnamefont {Glaser}}, \ and\ \bibinfo
		{author} {\bibfnamefont {D.}~\bibnamefont {Sugny}},\ }\bibfield  {title}
	{{\enquote {\bibinfo {title} {Linking the rotation of a
					rigid body to the {Schrödinger} equation: {The} quantum tennis racket effect
					and beyond},}\ }}\href {\doibase 10.1038/s41598-017-04174-x} {\bibfield
		{journal} {\bibinfo  {journal} {Scientific Reports}\ }\textbf {\bibinfo
			{volume} {7}},\ \bibinfo {pages} {1--8} (\bibinfo {year} {2017})}\BibitemShut
	{NoStop}%
	\bibitem [{\citenamefont {Gerving}\ \emph {et~al.}(2012)\citenamefont
		{Gerving}, \citenamefont {Hoang}, \citenamefont {Land}, \citenamefont
		{Anquez}, \citenamefont {Hamley},\ and\ \citenamefont
		{Chapman}}]{gerving2012non}%
	\BibitemOpen
	\bibfield  {author} {\bibinfo {author} {\bibfnamefont {C.~S.}\ \bibnamefont
			{Gerving}}, \bibinfo {author} {\bibfnamefont {T.~M.}\ \bibnamefont {Hoang}},
		\bibinfo {author} {\bibfnamefont {B.J.}\ \bibnamefont {Land}}, \bibinfo
		{author} {\bibfnamefont {M.}~\bibnamefont {Anquez}}, \bibinfo {author}
		{\bibfnamefont {C.~D.}\ \bibnamefont {Hamley}}, \ and\ \bibinfo {author}
		{\bibfnamefont {M.~S.}\ \bibnamefont {Chapman}},\ }\bibfield  {title}
	{\enquote {\bibinfo {title} {Non-equilibrium dynamics of an unstable quantum
				pendulum explored in a spin-1 bose--einstein condensate},}\ }\href@noop {}
	{\bibfield  {journal} {\bibinfo  {journal} {Nat. Commun.}\ }\textbf {\bibinfo
			{volume} {3}},\ \bibinfo {pages} {1--7} (\bibinfo {year} {2012})}\BibitemShut
	{NoStop}%
	\bibitem [{\citenamefont {Chuchem}\ \emph {et~al.}(2010)\citenamefont
		{Chuchem}, \citenamefont {Smith-Mannschott}, \citenamefont {Hiller},
		\citenamefont {Kottos}, \citenamefont {Vardi},\ and\ \citenamefont
		{Cohen}}]{chuchem2010quantum}%
	\BibitemOpen
	\bibfield  {author} {\bibinfo {author} {\bibfnamefont {Maya}\ \bibnamefont
			{Chuchem}}, \bibinfo {author} {\bibfnamefont {Katrina}\ \bibnamefont
			{Smith-Mannschott}}, \bibinfo {author} {\bibfnamefont {Moritz}\ \bibnamefont
			{Hiller}}, \bibinfo {author} {\bibfnamefont {Tsampikos}\ \bibnamefont
			{Kottos}}, \bibinfo {author} {\bibfnamefont {Amichay}\ \bibnamefont {Vardi}},
		\ and\ \bibinfo {author} {\bibfnamefont {Doron}\ \bibnamefont {Cohen}},\
	}\bibfield  {title} {\enquote {\bibinfo {title} {Quantum dynamics in the
				bosonic josephson junction},}\ }\href@noop {} {\bibfield  {journal} {\bibinfo
			{journal} {Phys. Rev. A}\ }\textbf {\bibinfo {volume} {82}},\ \bibinfo
		{pages} {053617} (\bibinfo {year} {2010})}\BibitemShut {NoStop}%
	\bibitem [{\citenamefont {Hamraoui}\ \emph {et~al.}(2018)\citenamefont
		{Hamraoui}, \citenamefont {Van~Damme}, \citenamefont {Marde\ifmmode
			\check{s}\else \v{s}\fi{}i\ifmmode~\acute{c}\else \'{c}\fi{}},\ and\
		\citenamefont {Sugny}}]{hamraoui2018}%
	\BibitemOpen
	\bibfield  {author} {\bibinfo {author} {\bibfnamefont {K.}~\bibnamefont
			{Hamraoui}}, \bibinfo {author} {\bibfnamefont {L.}~\bibnamefont {Van~Damme}},
		\bibinfo {author} {\bibfnamefont {P.}~\bibnamefont {Marde\ifmmode
				\check{s}\else \v{s}\fi{}i\ifmmode~\acute{c}\else \'{c}\fi{}}}, \ and\
		\bibinfo {author} {\bibfnamefont {D.}~\bibnamefont {Sugny}},\ }\bibfield
	{title} {\enquote {\bibinfo {title} {Classical and quantum rotation numbers
				of asymmetric-top molecules},}\ }\href {\doibase 10.1103/PhysRevA.97.032118}
	{\bibfield  {journal} {\bibinfo  {journal} {Phys. Rev. A}\ }\textbf {\bibinfo
			{volume} {97}},\ \bibinfo {pages} {032118} (\bibinfo {year}
		{2018})}\BibitemShut {NoStop}%
	\bibitem [{\citenamefont {Ashbaugh}\ \emph {et~al.}(1991)\citenamefont
		{Ashbaugh}, \citenamefont {Chicone},\ and\ \citenamefont
		{Cushman}}]{ashbaugh_twisting_1991}%
	\BibitemOpen
	\bibfield  {author} {\bibinfo {author} {\bibfnamefont {Mark~S.}\ \bibnamefont
			{Ashbaugh}}, \bibinfo {author} {\bibfnamefont {Carmen~C.}\ \bibnamefont
			{Chicone}}, \ and\ \bibinfo {author} {\bibfnamefont {Richard~H.}\
			\bibnamefont {Cushman}},\ }\bibfield  {title} {{\enquote
			{\bibinfo {title} {The twisting tennis racket},}\ }}\href {\doibase
		10.1007/BF01049489} {\bibfield  {journal} {\bibinfo  {journal} {Journal of
				Dynamics and Differential Equations}\ }\textbf {\bibinfo {volume} {3}},\
		\bibinfo {pages} {67--85} (\bibinfo {year} {1991})}\BibitemShut {NoStop}%
	\bibitem [{\citenamefont {Van~Damme}\ \emph {et~al.}(2017)\citenamefont
		{Van~Damme}, \citenamefont {Marde{\v{s}}i{\'c}},\ and\ \citenamefont
		{Sugny}}]{van2017tennis}%
	\BibitemOpen
	\bibfield  {author} {\bibinfo {author} {\bibfnamefont {L{\'e}o}\ \bibnamefont
			{Van~Damme}}, \bibinfo {author} {\bibfnamefont {Pavao}\ \bibnamefont
			{Marde{\v{s}}i{\'c}}}, \ and\ \bibinfo {author} {\bibfnamefont {Dominique}\
			\bibnamefont {Sugny}},\ }\bibfield  {title} {\enquote {\bibinfo {title} {The
				tennis racket effect in a three-dimensional rigid body},}\ }\href@noop {}
	{\bibfield  {journal} {\bibinfo  {journal} {Physica D}\ }\textbf {\bibinfo
			{volume} {338}},\ \bibinfo {pages} {17--25} (\bibinfo {year}
		{2017})}\BibitemShut {NoStop}%
	\bibitem [{sup()}]{supp}%
	\BibitemOpen
	\href@noop {} {\bibinfo  {journal} {See Supplementary Information}\
	}\BibitemShut {NoStop}%
	\bibitem [{\citenamefont {Landau}\ and\ \citenamefont
		{Lifshitz}(2013)}]{landau2013quantum}%
	\BibitemOpen
	\bibfield  {journal} {  }\bibfield  {author} {\bibinfo {author} {\bibfnamefont
			{Lev~Davidovich}\ \bibnamefont {Landau}}\ and\ \bibinfo {author}
		{\bibfnamefont {Evgenii~Mikhailovich}\ \bibnamefont {Lifshitz}},\ }\href@noop
	{} {\emph {\bibinfo {title} {Quantum mechanics: non-relativistic theory}}},\
	Vol.~\bibinfo {volume} {3}\ (\bibinfo  {publisher} {Elsevier},\ \bibinfo
	{year} {2013})\BibitemShut {NoStop}%
	\bibitem [{\citenamefont {Harter}\ and\ \citenamefont
		{Patterson}(1984)}]{harter1984rotational}%
	\BibitemOpen
	\bibfield  {author} {\bibinfo {author} {\bibfnamefont {William~G}\
			\bibnamefont {Harter}}\ and\ \bibinfo {author} {\bibfnamefont {Chris~W}\
			\bibnamefont {Patterson}},\ }\bibfield  {title} {\enquote {\bibinfo {title}
			{Rotational energy surfaces and high-j eigenvalue structure of polyatomic
				molecules},}\ }\href@noop {} {\bibfield  {journal} {\bibinfo  {journal} {J.
				Chem. Phys.}\ }\textbf {\bibinfo {volume} {80}},\ \bibinfo {pages}
		{4241--4261} (\bibinfo {year} {1984})}\BibitemShut {NoStop}%
	\bibitem [{\citenamefont {Rundle}\ \emph {et~al.}(2019)\citenamefont {Rundle},
		\citenamefont {Tilma}, \citenamefont {Samson}, \citenamefont {Dwyer},
		\citenamefont {Bishop},\ and\ \citenamefont {Everitt}}]{rundle2019general}%
	\BibitemOpen
	\bibfield  {author} {\bibinfo {author} {\bibfnamefont {Russell~P}\
			\bibnamefont {Rundle}}, \bibinfo {author} {\bibfnamefont {Todd}\ \bibnamefont
			{Tilma}}, \bibinfo {author} {\bibfnamefont {JH}~\bibnamefont {Samson}},
		\bibinfo {author} {\bibfnamefont {Vincent~M}\ \bibnamefont {Dwyer}}, \bibinfo
		{author} {\bibfnamefont {RF}~\bibnamefont {Bishop}}, \ and\ \bibinfo {author}
		{\bibfnamefont {Mark~J}\ \bibnamefont {Everitt}},\ }\bibfield  {title}
	{\enquote {\bibinfo {title} {General approach to quantum mechanics as a
				statistical theory},}\ }\href@noop {} {\bibfield  {journal} {\bibinfo
			{journal} {Phys. Rev. A}\ }\textbf {\bibinfo {volume} {99}},\ \bibinfo
		{pages} {012115} (\bibinfo {year} {2019})}\BibitemShut {NoStop}%
	\bibitem [{\citenamefont {Korobenko}\ \emph {et~al.}(2014)\citenamefont
		{Korobenko}, \citenamefont {Milner},\ and\ \citenamefont
		{Milner}}]{korobenko2014}%
	\BibitemOpen
	\bibfield  {author} {\bibinfo {author} {\bibfnamefont {Aleksey}\ \bibnamefont
			{Korobenko}}, \bibinfo {author} {\bibfnamefont {Alexander~A}\ \bibnamefont
			{Milner}}, \ and\ \bibinfo {author} {\bibfnamefont {Valery}\ \bibnamefont
			{Milner}},\ }\bibfield  {title} {\enquote {\bibinfo {title} {Direct
				observation, study, and control of molecular superrotors},}\ }\href@noop {}
	{\bibfield  {journal} {\bibinfo  {journal} {Phys. Rev. Lett.}\ }\textbf
		{\bibinfo {volume} {112}},\ \bibinfo {pages} {113004} (\bibinfo {year}
		{2014})}\BibitemShut {NoStop}%
	\bibitem [{\citenamefont {Tebbenjohanns}\ \emph {et~al.}(2020)\citenamefont
		{Tebbenjohanns}, \citenamefont {Frimmer}, \citenamefont {Jain}, \citenamefont
		{Windey},\ and\ \citenamefont {Novotny}}]{tebbenjohanns2020}%
	\BibitemOpen
	\bibfield  {author} {\bibinfo {author} {\bibfnamefont {Felix}\ \bibnamefont
			{Tebbenjohanns}}, \bibinfo {author} {\bibfnamefont {Martin}\ \bibnamefont
			{Frimmer}}, \bibinfo {author} {\bibfnamefont {Vijay}\ \bibnamefont {Jain}},
		\bibinfo {author} {\bibfnamefont {Dominik}\ \bibnamefont {Windey}}, \ and\
		\bibinfo {author} {\bibfnamefont {Lukas}\ \bibnamefont {Novotny}},\
	}\bibfield  {title} {\enquote {\bibinfo {title} {Motional sideband asymmetry
				of a nanoparticle optically levitated in free space},}\ }\href {\doibase
		10.1103/PhysRevLett.124.013603} {\bibfield  {journal} {\bibinfo  {journal}
			{Phys. Rev. Lett.}\ }\textbf {\bibinfo {volume} {124}},\ \bibinfo {pages}
		{013603} (\bibinfo {year} {2020})}\BibitemShut {NoStop}%
	\bibitem [{\citenamefont {Rider}\ \emph {et~al.}(2016)\citenamefont {Rider},
		\citenamefont {Moore}, \citenamefont {Blakemore}, \citenamefont {Louis},
		\citenamefont {Lu},\ and\ \citenamefont {Gratta}}]{rider2016search}%
	\BibitemOpen
	\bibfield  {author} {\bibinfo {author} {\bibfnamefont {Alexander~D}\
			\bibnamefont {Rider}}, \bibinfo {author} {\bibfnamefont {David~C}\
			\bibnamefont {Moore}}, \bibinfo {author} {\bibfnamefont {Charles~P}\
			\bibnamefont {Blakemore}}, \bibinfo {author} {\bibfnamefont {Maxime}\
			\bibnamefont {Louis}}, \bibinfo {author} {\bibfnamefont {Marie}\ \bibnamefont
			{Lu}}, \ and\ \bibinfo {author} {\bibfnamefont {Giorgio}\ \bibnamefont
			{Gratta}},\ }\bibfield  {title} {\enquote {\bibinfo {title} {Search for
				screened interactions associated with dark energy below the 100 $\mu$m length
				scale},}\ }\href@noop {} {\bibfield  {journal} {\bibinfo  {journal} {Phys.
				Rev. Lett.}\ }\textbf {\bibinfo {volume} {117}},\ \bibinfo {pages} {101101}
		(\bibinfo {year} {2016})}\BibitemShut {NoStop}%
	\bibitem [{\citenamefont {Karczmarek}\ \emph {et~al.}(1999)\citenamefont
		{Karczmarek}, \citenamefont {Wright}, \citenamefont {Corkum},\ and\
		\citenamefont {Ivanov}}]{karczmarek1999optical}%
	\BibitemOpen
	\bibfield  {author} {\bibinfo {author} {\bibfnamefont {Joanna}\ \bibnamefont
			{Karczmarek}}, \bibinfo {author} {\bibfnamefont {James}\ \bibnamefont
			{Wright}}, \bibinfo {author} {\bibfnamefont {Paul}\ \bibnamefont {Corkum}}, \
		and\ \bibinfo {author} {\bibfnamefont {Misha}\ \bibnamefont {Ivanov}},\
	}\bibfield  {title} {\enquote {\bibinfo {title} {Optical centrifuge for
				molecules},}\ }\href@noop {} {\bibfield  {journal} {\bibinfo  {journal}
			{Phys. Rev. Lett.}\ }\textbf {\bibinfo {volume} {82}},\ \bibinfo {pages}
		{3420} (\bibinfo {year} {1999})}\BibitemShut {NoStop}%
	\bibitem [{\citenamefont {Stickler}\ \emph
		{et~al.}(2016{\natexlab{b}})\citenamefont {Stickler}, \citenamefont
		{Papendell},\ and\ \citenamefont {Hornberger}}]{stickler2016spatio}%
	\BibitemOpen
	\bibfield  {author} {\bibinfo {author} {\bibfnamefont {Benjamin~A}\
			\bibnamefont {Stickler}}, \bibinfo {author} {\bibfnamefont {Birthe}\
			\bibnamefont {Papendell}}, \ and\ \bibinfo {author} {\bibfnamefont {Klaus}\
			\bibnamefont {Hornberger}},\ }\bibfield  {title} {\enquote {\bibinfo {title}
			{Spatio-orientational decoherence of nanoparticles},}\ }\href@noop {}
	{\bibfield  {journal} {\bibinfo  {journal} {Phys. Rev. A}\ }\textbf {\bibinfo
			{volume} {94}},\ \bibinfo {pages} {033828} (\bibinfo {year}
		{2016}{\natexlab{b}})}\BibitemShut {NoStop}%
\end{thebibliography}

\end{document}